\documentclass{article}

\usepackage{arxiv}
\usepackage[utf8]{inputenc} 
\usepackage[T1]{fontenc}    
\usepackage{hyperref}       
\usepackage{url}            
\usepackage{booktabs}       
\usepackage{amsfonts}       
\usepackage{nicefrac}       
\usepackage{microtype}      
\usepackage{graphicx}
\usepackage{amsmath}
\usepackage{booktabs}
\usepackage{multirow}
\usepackage{colortbl}
\usepackage{color}
\usepackage{caption}
\usepackage{array}
\usepackage{makecell}
\usepackage{subfigure}
\usepackage{dcolumn}
\usepackage{tabularx}
\usepackage{xpatch}
\xpatchcmd\citenum{\NAT@parfalse}{\NAT@partrue}{}{}
\usepackage{appendix}
\usepackage{titlesec}
\graphicspath{ {./images/} }

\title{On the Preprocessing of Physics-informed Neural Networks: How to Better Utilize Data in Fluid Mechanics}

\author{
  Shengfeng Xu\\
  Institute of Mechanics\\
  Chinese Academy of Sciences\\
  Beijing 100190, China\\
  Integrative Sciences and Engineering Programme\\
  NUS Graduate School\\
  National University of Singapore\\
  Singapore 119077, Singapore\\
   \And
  Chang Yan\\
  Institute of Mechanics\\
  Chinese Academy of Sciences\\
  Beijing 100190, China\\
  School of Future Technology\\
  University of Chinese Academy of Sciences\\
  Beijing 100049, China\\
   \And
  Zhenxu Sun\thanks{Corresponding author}\\
  Institute of Mechanics\\
  Chinese Academy of Sciences\\
  Beijing 100190, China\\
  \texttt{sunzhenxu@imech.ac.cn} \\
  \And
  Renfang Huang\\
  Institute of Mechanics\\
  Chinese Academy of Sciences\\
  Beijing 100190, China\\
  \And
  Dilong Gu\\
  Institute of Mechanics\\
  Chinese Academy of Sciences\\
  Beijing 100190, China\\
  \And
  Guowei Yang\\
  Institute of Mechanics\\
  Chinese Academy of Sciences\\
  Beijing 100190, China\\
}

\begin{document}
\maketitle
\begin{abstract}
Physics-Informed Neural Networks (PINNs) serve as a flexible alternative for tackling forward and inverse problems in differential equations, displaying impressive advancements in diverse areas of applied mathematics. Despite integrating both data and underlying physics to enrich the neural network's understanding, concerns regarding the effectiveness and practicality of PINNs persist. Over the past few years, extensive efforts in the current literature have been made to enhance this evolving method, by drawing inspiration from both machine learning algorithms and numerical methods. Despite notable progressions in PINNs algorithms, the important and fundamental field of data preprocessing remain unexplored, limiting the applications of PINNs especially in solving inverse problems. Therefore in this paper, a concise yet potent data preprocessing method focusing on data normalization was proposed. By applying a linear transformation to both the data and corresponding equations concurrently, the normalized PINNs approach was evaluated on the task of reconstructing flow fields in three turbulent cases. The results illustrate that by adhering to the data preprocessing procedure, PINNs can robustly achieve higher prediction accuracy for all flow quantities under different hyperparameter setups, without incurring extra computational cost, distinctly improving the utilization of limited training data. Though only verified in Navier-Stokes (NS) equations, this method holds potential for application to various other equations.
\end{abstract}
\section{\label{sec: INTRODUCTION}INTRODUCTION}
Machine learning has rapidly permeated numerous domains over the last two decades, encompassing daily life\cite{kiran2021deep,portugal2018use} and scientific research\cite{morgan2020opportunities,greener2022guide,brunton2020machine}, significantly changed our way of living and studying through the exploitation of vast datasets. Unlike pure data-driven approaches, Physics-informed Neural Networks is developed based on the concept of utilizing both data and underlying physical laws\cite{karniadakis2021physics,cuomo2022scientific}. Originally conceived by Lagaris\cite{lagaris1998artificial} and rekindled by Raissi\cite{raissi2019physics}, PINNs can now easily be deployed via popular machine learning framework such as TensorFlow and PyTorch. This method provides a flexible framework for solving both forward and inverse problems\cite{yang2021b,yu2022gradient,zhang2023enforcing} associated with differential equations, finding applications in diverse fields such as  heat transfer\cite{cai2021physics,jagtap2023coolpinns}, seismology\cite{ren2024seismicnet}, quantum physics\cite{zhou2021solving, jin2022physics} and fluid mechanics\cite{cai2021fluid,sharma2023review}.

Fluid mechanics, governed by the well-known Navier-Stokes (NS) equations and characterized by its nonlinear nature, stands out as a primary domain showcasing diverse applications of PINNs, ranging from laminar flows\cite{rao2020physics,biswas2023three} to turbulent flows\cite{jin2021nsfnets, hanrahan2023studying}. Impressive examples include pressure and velocity inference from concentration field in cylinder wake and artery\cite{raissi2020hidden}, pressure and velocity reconstruction from temperature fields over an espresso cup\cite{cai2021flow}, super-resolution and denoising of time resolved three-dimensional phase contrast magnetic resonance imaging (4D-Flow MRI)\cite{fathi2020super}, etc. Despite the adaptability of PINNs in addressing both forward and inverse problems, given the convenience of automatic differentiation\cite{baydin2018automatic} in deep learning frameworks, and there have been evidence of successful flow simulation of PINNs without any labeled data in the training set\cite{sun2020surrogate}. However, due to the unpredictable accuracy and prohibitive computational cost, the current nascent stage PINNs method is not comparable to traditional numerical algorithms in forward problems, leaving the superiority of PINNs mainly in inverse problems, where unknown flow features or undetermined parameters can be derived through abundant data or sparse data with ill-posed condition, which is also the focus of this paper.

A typical example of such inverse problems in fluid mechanics is flow field reconstruction, where incomplete or sparse data can be utilized by PINNs to reconstruct the unknown flow field\cite{xu2023practical}. Xu et al.\cite{xu2021explore} reconstructed the missing flow dynamics by treating the governing equations as a parameterized constraint. Wang et al. \cite{wang2024dynamic} reconstructed the wake field of wind turbine using sparse LiDAR data. Xu et al.\cite{xu2023spatiotemporal} reconstructed the cylinder wake and decaying turbulence using regularly distributed sparse data. Liu et al.\cite{liu2024high} reconstructed three-dimensional turbulent combustion with high-resolution using synthetic sparse data.

Despite the successful reconstructions mentioned above, the evolving method of PINNs still holds the potential to achieve tasks with increased accuracy and efficiency. In recent years, significant efforts have been dedicated to improving the trainability of PINNs, such as domain decomposition\cite{jagtap2020conservative, hu2023augmented}, parallel computing\cite{meng2020ppinn, xu2023spatiotemporal}, adaptive sampling by respecting the causality\cite{wang2024respecting}, hybrid numerical scheme\cite{chiu2022can}, positional encoding for complex geometry\cite{costabal2024delta}, etc. While many of these improvements are developed by taking reference from traditional numerical methods, indeed PINNs can also be enhanced by adopting "tricks" from mainstream machine learning algorithms, such as adaptive activation function\cite{jagtap2020locally,jagtap2020adaptive}, adaptive weighting\cite{mcclenny2023self}, transfer learning\cite{xu2023transfer}, etc. Among all these "tricks", one area that has received little attention in the realm of PINNs is data preprocessing.

As a fundamental technique in machine learning, data preprocessing methods such as data cleaning, data transformation, and data normalization\cite{garcia2015data} have non-negligible effect on the performance of supervised learning\cite{kotsiantis2006data} and unsupervised learning\cite{garcia2016big}. Among these methods, data normalization stands out as the most widely employed in various machine learning algorithms, aiming to scale all features to comparable level for more accurate learning. While analogous strategies, like incorporating non-dimensional equations, have been utilized in PINNs, and there has also been a worthwhile attempt made by introducing a hidden normalization layer in the neural networks architecture\cite{raissi2020hidden} (which would also be discussed in this paper). The effectiveness of data normalization in PINNs remains underexplored. Therefore in this paper, a data preprocessing pipeline focusing on normalization was proposed and evaluated on three test cases involving the reconstruction of unsteady flows with sparse data.

This article is organized as follows: Section \ref{sec: METHODOLOGY} provides a detailed description of PINNs and the proposed normalization method on PINNs. In Section \ref{sec: NUMERICAL DATASET USED FOR TRAINING AND VALIDATION}, the three sparse datasets used in this study are introduced. Section \ref{sec: RESULTS} demonstrates the superiority of the proposed preprocessing pipeline on three distinct problems compared to existing methods. Section \ref{sec: ANALYSIS} analyzes why the proposed preprocessing method presents superior performance, and Section \ref{sec: CONCLUSION AND DISCUSSION} concludes the paper with a brief summary and discussion.

\section{\label{sec: METHODOLOGY}METHODOLOGY}
\subsection{Physics informed neural networks}
\begin{figure*}[!htbp]
 \centering
 \includegraphics[width=0.95\textwidth]{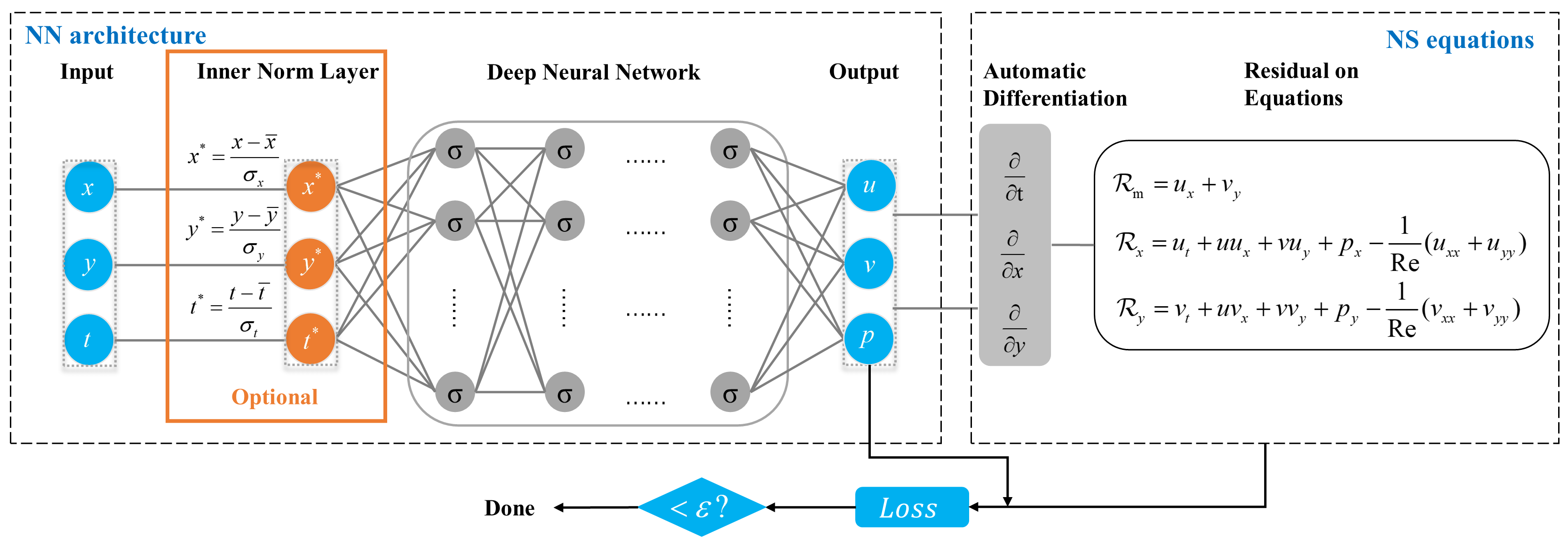}
  \caption{Physics-informed neural networks for solving non-dimensional NS equations. The loss function consists of both the data loss and the residual error of the conservation equations. $\sigma$ denotes the activation function. The inner normalization layer is an optional hidden layer that normalize the input features with z-score normalization, using mean and standard deviation values from the training set.}
 \label{fig:pinns}
\end{figure*}

PINNs is a mesh-free approach for solving ordinary differential equations\cite{de2024physics} and partial differential equations\cite{mishra2023estimates}. In addition to the conventional function of neural networks in data fitting, PINNs can further incorporate physical constraints: By configuring the independent variables as network input and the dependent variables as network output, the corresponding derivatives can be computed numerically by backtracking the computational graph. These derivatives are then combined with variables to construct the residuals of the differential equations. Subsequently, by integrating these residuals into the loss function, the physical laws can be imposed in a soft manner, as illustrated in Fig.\ref{fig:pinns}.

Given a neural network $\tilde u({\bf{x}};\theta)$, where $\bf{x}$ means the network input and $\theta$ means the undetermined weights and biases in the network, the composite loss function containing the data loss term and equation loss term can be written as 
\begin{equation}
\zeta(\theta)   = {\zeta_{d}}(\theta) + \lambda_{r}  \times {\zeta_{r}}(\theta)
 \label{eq:loss}
\end{equation}
where
\begin{equation}
{\zeta_{d}}(\theta) = \frac{1}{{{N_d}}}\sum\limits_{i = 1}^{{N_d}} {{{\left\| {\tilde u({\bf{x}}_d^i;\theta) - {u_d^i}} \right\|}^2}}
 \label{eq:loss_data}
\end{equation}
\begin{equation}
{\zeta_{r}}(\theta) = \frac{1}{{{N_r}}}\sum\limits_{i = 1}^{Nr} {{{\left\| {{\bf{R}}(\tilde u({\bf{x}}_r^i;\theta))} \right\|}^2}} 
 \label{eq:loss_residual}
\end{equation}

where $\{ {\bf{x}}_d^i, u_d^i\} _{i = 1}^{{N_d}}$ is the training dataset with $N_d$ data points (where initial condition, boundary condition, and extra data information are included) and $\{ {\bf{x}}_r^i\} _{i = 1}^{{N_r}}$ is the residual set with $N_r$ residual points (where residuals of equations are evaluated). $\lambda_{r}$ is a tunable hyper-parameter that allows the flexibility of assigning a different learning rate to each loss term and can also be a dynamic value that evolves with iterations to accelerate convergence\cite{maddu2022inverse}. Normally, $\lambda_{r}$ is set to 1. ${\bf{R}}(\tilde u({\bf{x}};\theta))$ is the residual vector of underlying differential equations. For 2-dimensional incompressible flow, ${\bf{R}}(\tilde u({\bf{x}};\theta))$ consists of ${\cal R}_{m}$, ${\cal R}_{x}$, and ${\cal R}_{y}$, whose non-dimensional form can be written as
 \begin{equation}
{{\cal R}_{m}} = {u_x} + {v_y}
 \label{eq:continuity}
\end{equation}
\begin{equation}
{{\cal R}_x} = {u_t} + u{u_x} + v{u_y} + {p_x} - \frac{1}{{{\mathop{\rm Re}\nolimits} }}({u_{xx}} + {u_{yy}})
 \label{eq:momentum_X}
\end{equation}
\begin{equation}
{{\cal R}_y} = {v_t} + u{v_x} + v{v_y} + {p_y} - \frac{1}{{{\mathop{\rm Re}\nolimits} }}({v_{xx}} + {v_{yy}})
 \label{eq:momentum_y}
\end{equation}

By minimizing Eq.\ref{eq:loss}, PINNs can be made possible not only fitting the real data in the training set, but also satisfying the underlying physical principles, therefore capable of predicting unknown information of the flow field.
\subsection{\label{subsec:normalization}Normalization}
\begin{figure*}[!htbp]
\centering
 \includegraphics[width=0.95\textwidth]{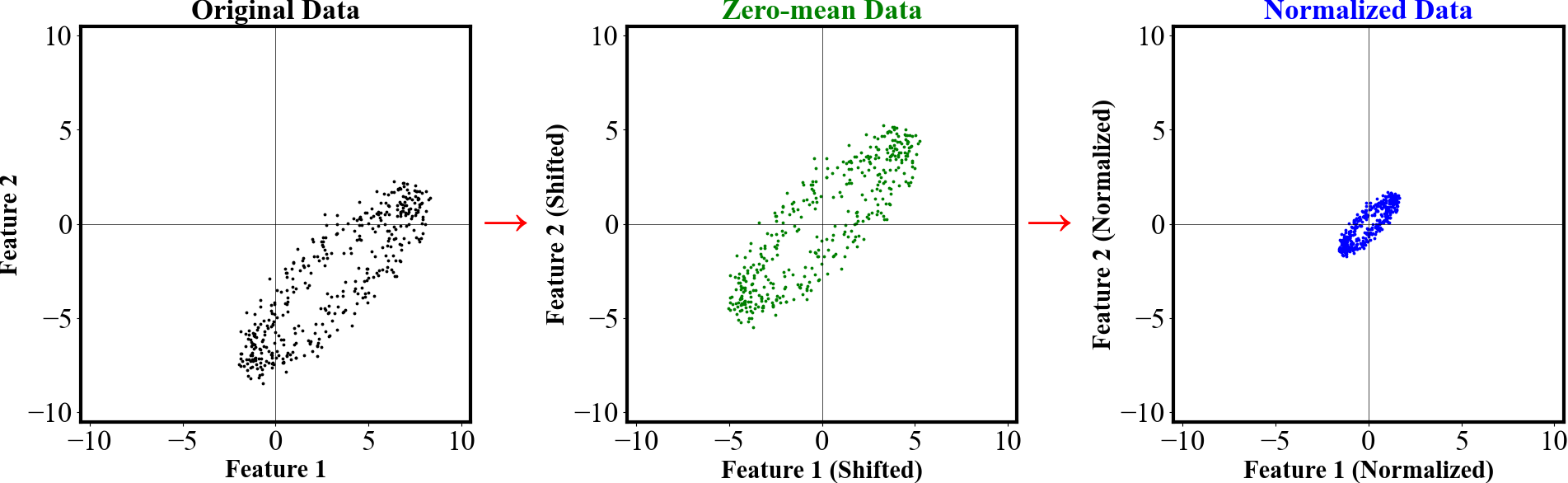}
  \caption{Two-dimensional illustration of Z-score normalization. \emph{Left}: original data. \emph{Middle}: the data has zero mean by subtracting the mean value on each dimension. \emph{Right}: the data further has standard deviation of 1 by dividing the standard deviation of original data.}
 \label{fig:normalization_demo}
\end{figure*}

\begin{figure*}[!htbp]
\centering
\includegraphics[width=0.95\textwidth]{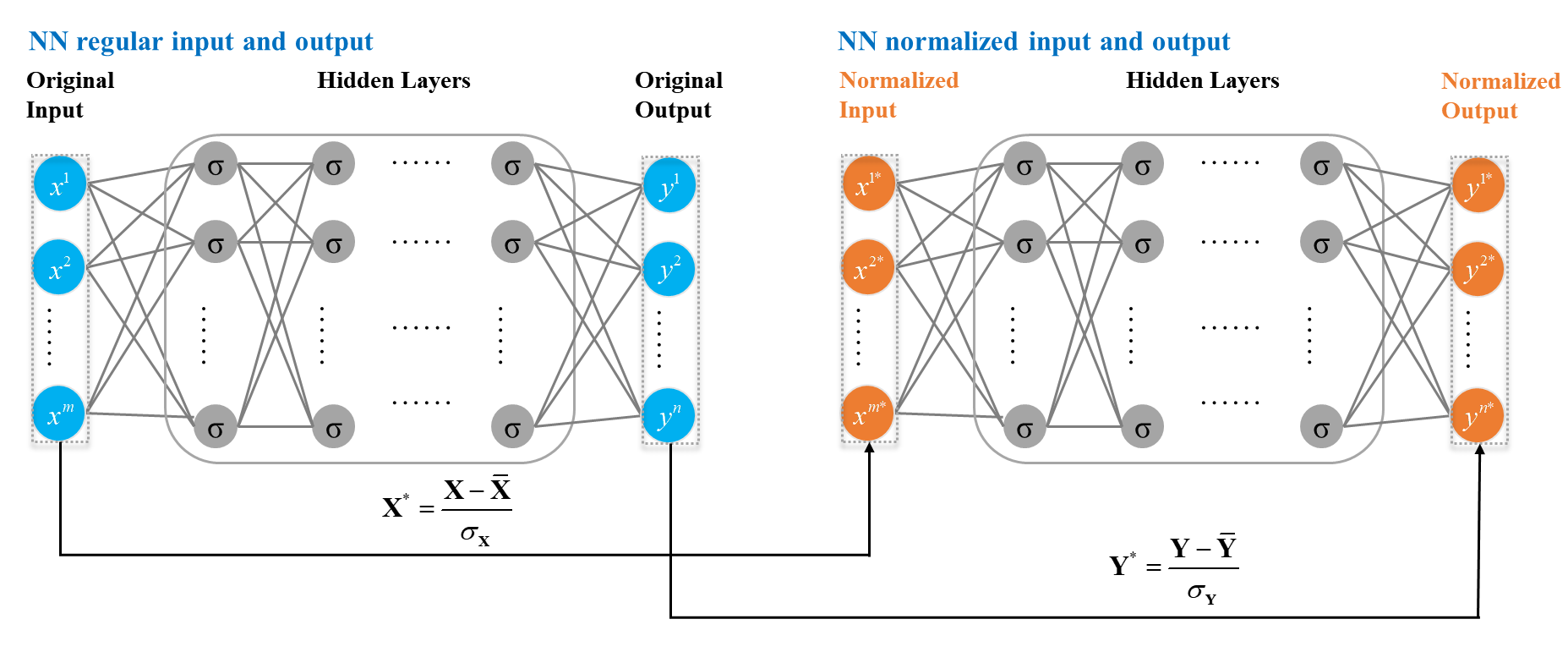}
\caption{Neural network structure comparison diagram. \emph{Left}: A neural network built on the original input-output feature mapping relationship. \emph{Right}: A neural network built on the normalized input-output feature mapping relationship.}
\label{fig:regular_normal_NN}
\end{figure*}

Data normalization includes Min-Max normalization, Z-score normalization, and Decimal Scaling normalization\cite{garcia2015data}, where simple linear transformation is performed on both input features and output features. Taking Z-score normalization as an example. The features are normalized by first subtracting the mean values and then dividing the standard deviations, as shown in Eq. \ref{eq:normalization}.
\begin{equation}
{{\bf{X}}^*} = \frac{{\bf{X} - {\bf{\bar X}}}}{{{\sigma _{\bf{X}}}}}
    \label{eq:normalization}
\end{equation}
By conducting Z-score normalization, the features are transformed to have a mean of 0 and a standard deviation of 1, as depicted in Fig.\ref{fig:normalization_demo}. After sufficient training, when the neural network learns the mapping relationship of normalized input-output features, the original input-output mapping relationship can be restored by denormalization, as shown in Eq.\ref{eq:denormalization_x}.
\begin{equation}
{\bf{X}} = {{\bf{X}}^*} \odot {\sigma _{\bf{X}}} + {\bf{\bar X}}
    \label{eq:denormalization_x}
\end{equation}
Where $\odot$ denotes element-wise product. Since Z-score normalization is the most commonly employed method and shares the same mathematical foundation with other normalization techniques, this paper exclusively concentrates on Z-score normalization, hereafter referred to simply as "normalization."
As illustrated in Fig. \ref{fig:normalization_demo}, normalization aims to scale neural network features to a comparable level, ensuring uniform contribution from each feature to the model's learning process. Consequently, it depicts the latent relationships in the data more accurately and leads to more precise learning outcomes. Therefore, in practical applications of neural networks, mapping relationships are typically established based on normalized input and output features, as depicted in Fig. \ref{fig:regular_normal_NN}, rather than the original features.

\begin{figure*}[!htbp]
\centering
\includegraphics[width=0.95\linewidth]{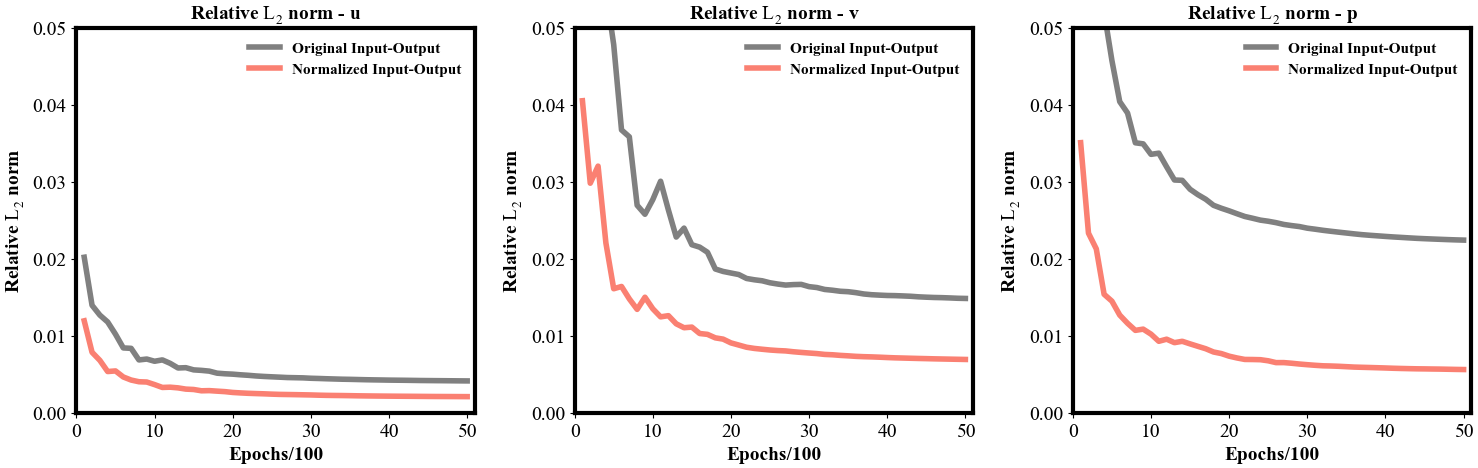}
\caption{Comparison of network fitting results. The relative $L_{2}$ norms of \emph{Left}: stream-wise velocity $u$, \emph{Middle}: transverse velocity $v$, \emph{Right}: pressure $p$ between benchmark flow data and fitted flow data using original input-output pairs and normalized input-output pairs.}
\label{fig:pureNN}
\end{figure*}

The efficacy of normalization can be clearly demonstrated through a simple fitting task using a flow dataset\footnote{The cylinder wake at $\rm{Re}=3900$, detailed in Section \ref{sec: NUMERICAL DATASET USED FOR TRAINING AND VALIDATION}}. The spatiotemporal coordinates of the flow field $\{x, y, t\}$ serve as the input, while the corresponding flow quantities $\{u, v, p\}$ are the output. The entire flow dataset is used as the training set for two different neural networks depicted in Fig. \ref{fig:regular_normal_NN} to perform data fitting (pure data fitting, without any physics or equations). The relative $L_2$ norm, as defined in Eq. \ref{eq:relativeL2}, is employed to quantify and compare the fitting accuracy.
\begin{equation}
{R_{L_{2}}} = \frac{{\left\| {\hat U - U} \right\|}}{{\left\| U \right\|}}
    \label{eq:relativeL2}
\end{equation}
Where ${\left\| U \right\|}$ denotes the $L_2$ norm of training data and ${\left\| {\hat U - U} \right\|}$ denotes the $L_2$ norm of the deviation between network prediction and training data. A lower value of ${R_{L_{2}}}$ indicates a more precise prediction/fitting. As presented in Fig.\ref{fig:pureNN}, the fitting of the training data is distinctly more accurate when adopting normalized input-output features, indicating the importance of applying normalization technique when training a neural network.
\subsection{Normalization on physics informed neural networks}
\subsubsection{Inner normalization layer}
Based on the above analysis and result, it is natural to try to adopt normalized features to achieve better prediction accuracy when training PINNs. However, in the case of PINNs, the features cannot be as conveniently normalized as in classic pure data-driven neural networks because each input and output feature has its own physical meaning. Applying normalization directly to PINNs would only result in inconsistencies between data and underlying equations. 

Given the importance of normalization, a previous attempt, presented in some popular open-source repositories\cite{raissi2020hidden}, involves incorporating a hidden layer, referred to as the inner normalization layer, immediately following the input layer. This inner normalization layer, depicted in the optional orange box in Fig.\ref{fig:pinns}, normalizes the input data using the mean and standard deviation values computed from the entire training dataset according to Eq.\ref{eq:normalization}. Since the normalization is conducted in the hidden layer, the input and output features remain unchanged, thus eliminating the need for a denormalization step. However, this approach solely normalizes the input data within the neural network, leaving the output data in their original state.
\subsubsection{Data normalization and equation transformation}
\begin{figure*}[!htbp]
\centering
\includegraphics[width=1.0\textwidth]{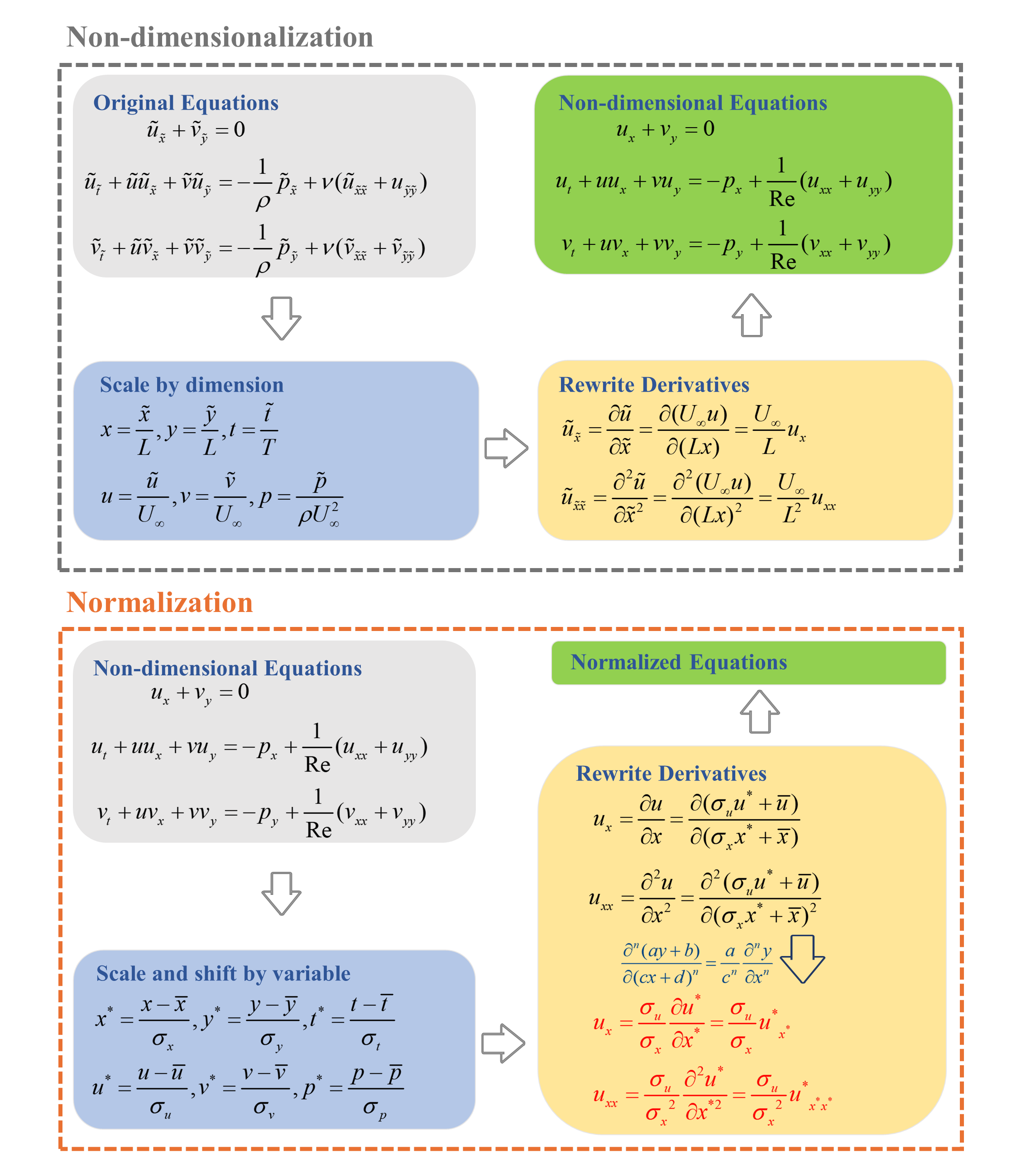}
\caption{Analogy diagram of the non-dimensionalization and normalization of NS equations. Top:  the process of non-dimensionalization of NS equations. Bottom:  the process of normalization of NS equations.}
\label{fig:NS_analogy}
\end{figure*}

Upon deeper consideration of data normalization, it becomes evident that the essence of normalization involves separate linear transformations for each variable, and the significant issue arises when these transformations result in a loss of coupling between the data and their respective equations. Interestingly, coupled linear transformations of both data and equations have long been established in the field of fluid mechanics, specifically in the non-dimensionalization of the Navier-Stokes (NS) equations\cite{patankar2018numerical}. As shown in the top box of Fig.\ref{fig:NS_analogy}, non-dimensionalization is implemented by scaling variables according to feature measurements in selected dimensions. To rectify this scaling, a subsequent rewriting in derivatives is required, resulting in a corresponding linear transformation of the equations. 
Similarly, normalization is achieved by scaling variables according to their statistical features rather than their dimensions. Unlike non-dimensionalization, normalization involves shifting and customized scaling for each variable. Thus, in this sense, normalization represents a more general process compared to non-dimensionalization. Therefore, in order to rectify normalization of NS equations, an analogous approach can be made by also first shifting and scaling variables and then rewriting the derivatives, as shown in the bottom box of Fig.\ref{fig:NS_analogy}. The details of normalization are as follows:

Considering all dependent variables and independent variables $\Psi  = {[x,y,t,u,v,p]^T}$ in the non-dimensional equations given by Eq.\ref{eq:continuity}-\ref{eq:momentum_y}, normalization can be achieved using mean values $\bar \Psi  = {[\bar x,\bar y,\bar t,\bar u,\bar v,\bar p]^T}$ and standard deviation values ${\sigma _\Psi } = {[{\sigma _x},{\sigma _y},{\sigma _t},{\sigma _u},{\sigma _v},{\sigma _p}]^T}$ computed from the training dataset. By applying Eq.\ref{eq:normalization}, the normalized variables ${\Psi ^*} = {[{x^*},{y^*},{t^*},{u^*},{v^*},{p^*}]^T}$ can be obtained, resulting in normalized training data with a mean of zero and a standard deviation of one for each variable. Subsequently, utilizing a simple calculus technique outlined in Eq.\ref{eq:calculus_trick}, Eq.\ref{eq:continuity}-\ref{eq:momentum_y} can be transformed into Eq.\ref{eq:trans_continuity}-\ref{eq:trans_momentum_y}.
\begin{equation}
    \frac{{{\partial ^n}(ay + b)}}{{\partial {{(x + d)}^n}}} = \frac{{a{\partial ^n}y}}{{{c^n}\partial {x^n}}}
    \label{eq:calculus_trick}
\end{equation}
 \begin{equation}
{{\cal R}_{m}}^* = \frac{{{\sigma _u}}}{{{\sigma _x}}}{u^*}_{{x^*}} + \frac{{{\sigma _v}}}{{{\sigma _y}}}{v^*}_{{y^*}}
 \label{eq:trans_continuity}
\end{equation}
\begin{equation}
\begin{split}
{{\cal R}_x}^* = \frac{{{\sigma _u}}}{{{\sigma _t}}}{u^*}_{{t^*}} + \hat u\frac{{{\sigma _u}}}{{{\sigma _x}}}{u^*}_{{x^*}} + \hat v\frac{{{\sigma _u}}}{{{\sigma _y}}}{u^*}_{{y^*}} + \frac{{{\sigma _p}}}{{{\sigma _x}}}{p^*}_{{x^*}}\\
- \frac{1}{{{\mathop{\rm Re}\nolimits} }}(\frac{{{\sigma _u}}}{{{\sigma _x}^2}}{u^*}_{{x^*}{x^*}} + \frac{{{\sigma _u}}}{{{\sigma _y}^2}}{u^*}_{{y^*}{y^*}})
\end{split}
 \label{eq:trans_momentum_X}
\end{equation}
\begin{equation}
\begin{split}
{{\cal R}_y}^* = \frac{{{\sigma _v}}}{{{\sigma _t}}}{v^*}_{{t^*}} + \hat u\frac{{{\sigma _v}}}{{{\sigma _x}}}{v^*}_{{x^*}} + \hat v\frac{{{\sigma _v}}}{{{\sigma _y}}}{v^*}_{{y^*}} + \frac{{{\sigma _p}}}{{{\sigma _y}}}{p^*}_{{y^*}}\\
- \frac{1}{{{\mathop{\rm Re}\nolimits} }}(\frac{{{\sigma _v}}}{{{\sigma _x}^2}}{v^*}_{{x^*}{x^*}} + \frac{{{\sigma _v}}}{{{\sigma _y}^2}}{v^*}_{{y^*}{y^*}})
\end{split}
 \label{eq:trans_momentum_y}
\end{equation}

\begin{figure*}[!htbp]
 \centering
 \includegraphics[width=0.9\textwidth]{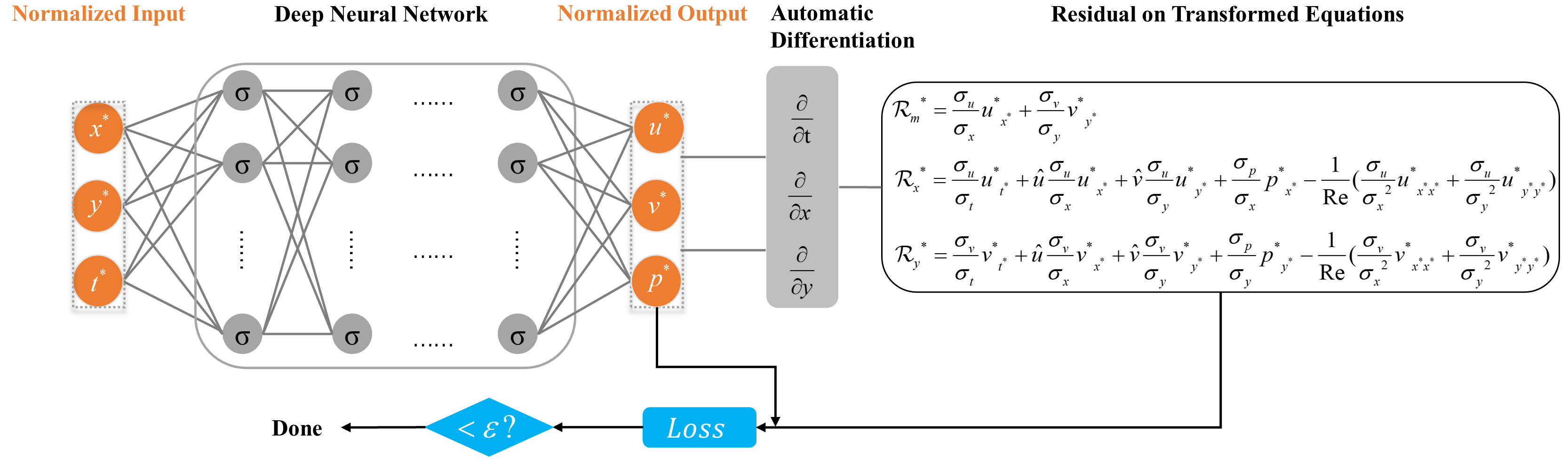}
  \caption{Normalized PINNs for solving non-dimensional equations. Both input and output variables are normalized using mean values and standard deviation values. The loss function comprises both the normalized data loss and the residual error of the transformed equations.}
 \label{fig:normalized_pinns}
\end{figure*}

Where $\hat u = ({\sigma _u}{u^*} + \bar u)$ and $\hat v = ({\sigma _v}{v^*} + \bar v)$. By simultaneously normalizing the data and transforming the equations, the adapted neural networks can be directly constructed using normalized variables. This approach, illustrated in Fig.\ref{fig:normalized_pinns}, facilitates the learning process for PINNs. Notably, the normalization process does not incur extra computational cost. Upon completion of training, the original variables can be restored through denormalization, as written in Eq.\ref{eq:denormalization_x}.
\section{\label{sec: NUMERICAL DATASET USED FOR TRAINING AND VALIDATION}NUMERICAL DATASET USED FOR TRAINING AND VALIDATION}
In this paper, three open source turbulent flow cases at different Reynolds numbers ($\rm{Re}$) were examined to evaluate the performance of the proposed preprocessing pipeline in reconstructing flow fields, i.e. the wake flow past a 2-dimensional (2D) circular cylinder at $\rm{Re}=3900$ (simulated with $k-\epsilon$ model), the wake flow past a 2D circular cylinder at $\rm{Re}=10000$\cite{yan2023exploring} (simulated with $k - \omega\,sst$ model), and 2D decaying turbulence at $\rm{Re}=2000$\cite{xu2023spatiotemporal} (simulated with pseudo-spectral method\cite{Lauber:2D-Turbulence-Python}).

\begin{figure*}[!htbp]
  \centering
  \includegraphics[width=0.85\textwidth]{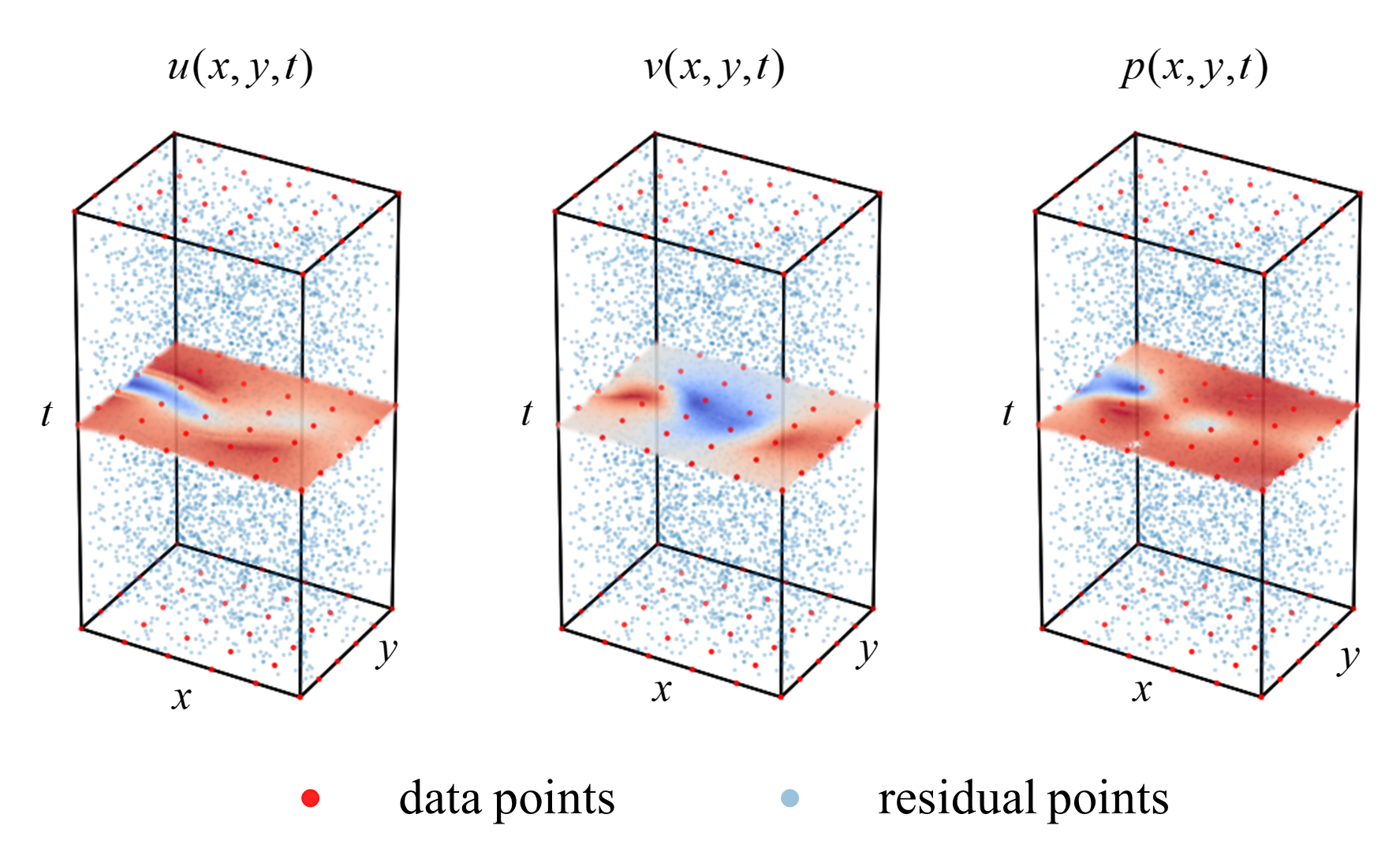}
  \caption{Schematic of sparse reconstruction training set for the wake flow of 2D cylinder. \emph{Left}: stream-wise velocity $u$. \emph{Middle}: transverse velocity $v$. \emph{Right}: pressure $p$. 36 sparsely distributed data points from 100 snapshots (only 3 snapshots are plotted) covering a period of 42.9 seconds are taken as training dataset, along with 1000000 residual points sampled through Latin Hypercube Sampling.}
  \label{fig:data&collocation_points}
\end{figure*}

The data used in flow field reconstruction is typically sparse in spatial domain and dense in temporal domain, simulating the realistic experiment setup. Taking the cylinder wake flow at Reynolds number $\rm{Re} = 3900$ case as an example, the sparse flow data ${u,v,p}$ is sampled from 36 sparsely distributed points covering a period of 42.9 seconds, as shown in Fig.\ref{fig:data&collocation_points}. The total number of data points $N_d$ is 3600, with 36 data points in each of the 100 snapshots. The overall number of residual points $N_r$ is 1000000, sampled through Latin Hypercube Sampling (LHS) method in the normalized spatio-temporal domain. The training sets for all three cases are similar, as presented in Table.\ref{tab: cfd_dataset}.
\begin{table}
\centering
\captionsetup{font={color=black}}
\caption{Training set information for flow reconstruction cases.}
\begin{tabularx}{1\textwidth}{c c c c c c}
\hline\hline
Case & $\rm{Re}$ & Data points & \makecell{Snapshots/\\Duration}& Residual points & \makecell{Domain\\$x\times y$} \\
\midrule
\multirow{2}{*}{cylinder wake} & 3900 & 3600 & 100/42.9s & 1000000 & $[1.000,5.000]\times[-2.000,2.000]$\\
& 10000 & 3600 & 100/2.86s & 1000000 & $[1.000,5.000]\times[-2.000,2.000]$\\
\midrule
decaying turbulence & 2000 & 3600 & 100/1.3s &1000000 & $[0.50\pi ,1.50\pi ]\times[0.50\pi ,1.50\pi ]$\\
\hline\hline
\end{tabularx}
\label{tab: cfd_dataset}
\end{table}
More detailed information and verification about these three cases can be referenced in literature \cite{xu2023spatiotemporal} and \cite{yan2023exploring}. The original dataset can be found in \href{https://github.com/Shengfeng233/PINN-MPI}{PINN-MPI} and \href{https://github.com/Panda000001/PINN-POD}{PINN-POD}.
\section{\label{sec: RESULTS}RESULTS}
In this section, three methods are evaluated and compared—namely, the commonly adopted non-dimensionalization method (referred to as Baseline), the existing inner normalization layer method (InnerNorm), and our proposed normalization method (Normalization)—in the context of three flow field reconstruction tasks with sparsely distributed flow data. Consistent with quantification methods commonly used in recent literature, the relative $L_2$ norm, as defined in Eq.\ref{eq:relativeL2}, is adopted to quantify the prediction error between the network output and the original data. More detailed qualitative and quantitative comparisons are provided in Appendix.\ref{Appendix:A} and Appendix.\ref{Appendix:B}.

Given that the performance of PINNs varies with different hyperparameter setups, to present a robust and comprehensive comparison, this paper evaluated the performance of the above three methods under different batch sizes, learning rate schedules, and parameter sizes for each case. According to recent literature\cite{raissi2020hidden,hao2023pinnacle}, these three important hyperparameters are sensitively related to the prediction accuracy and should be carefully manipulated when training PINNs. For all test cases, the optimizer used was Adam\cite{kingma2014adam}, the activation function was tanh, and the number of training epochs was 3000. Although the specific training time varied with different hyperparameters, the training times for the Baseline, InnerNorm, and Normalization methods were essentially the same under identical hyperparameter setups. With a batch size of 8192, a learning rate schedule of exponential decay, and a parameter size of 9731, the training time for all three methods on an NVIDIA Tesla V100-SXM2 GPU was approximately 4.8 seconds per epoch, indicating that normalization does not incur additional computational cost.
\subsection{\label{subsec: Batch size}Batch size}
\begin{figure*}[!htbp]
  \centering
\subfigure[Case 1: Cylinder wake at $\rm{Re}=3900$]{\label{fig:batch_size_compare_aveRe3900}
\includegraphics[width=0.95\linewidth]{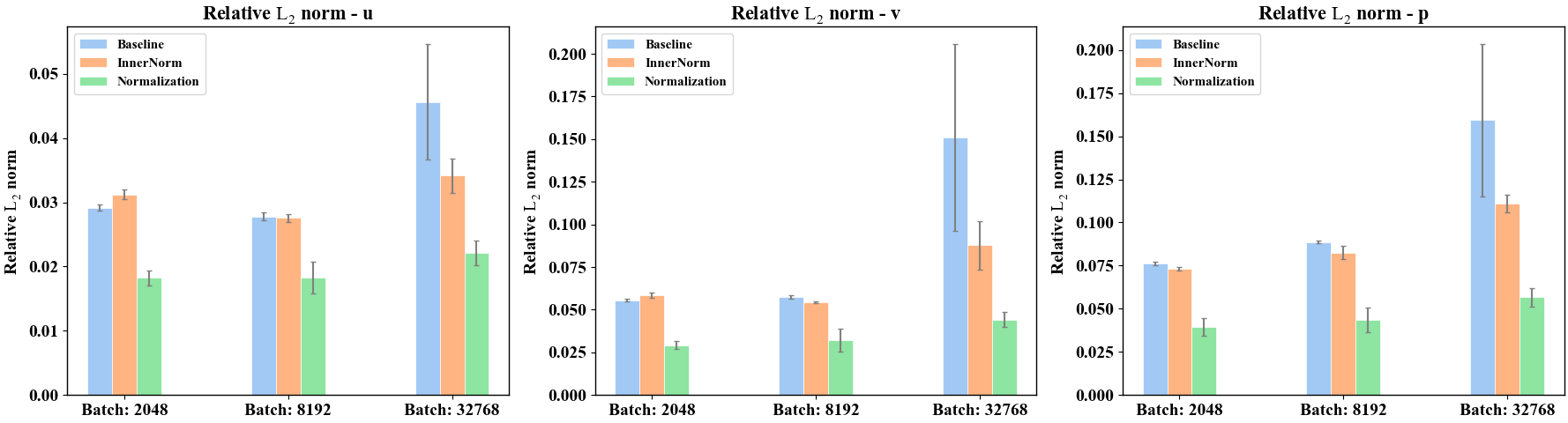}}
\hfill
\subfigure[Case 2: Cylinder wake at $\rm{Re}=10000$]{\label{fig:batch_size_compare_aveRe10000}
\includegraphics[width=0.95\linewidth]{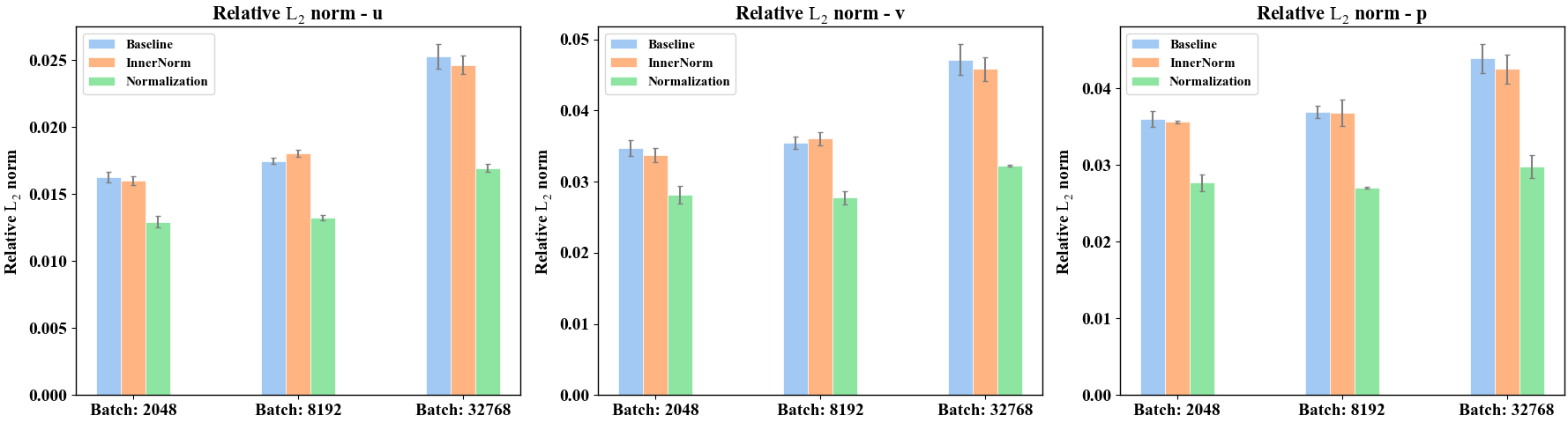}}
\hfill
\subfigure[Case 3: Decaying turbulence at $\rm{Re}=2000$]{\label{fig:batch_size_compare_aveRe2000}
\includegraphics[width=0.95\linewidth]{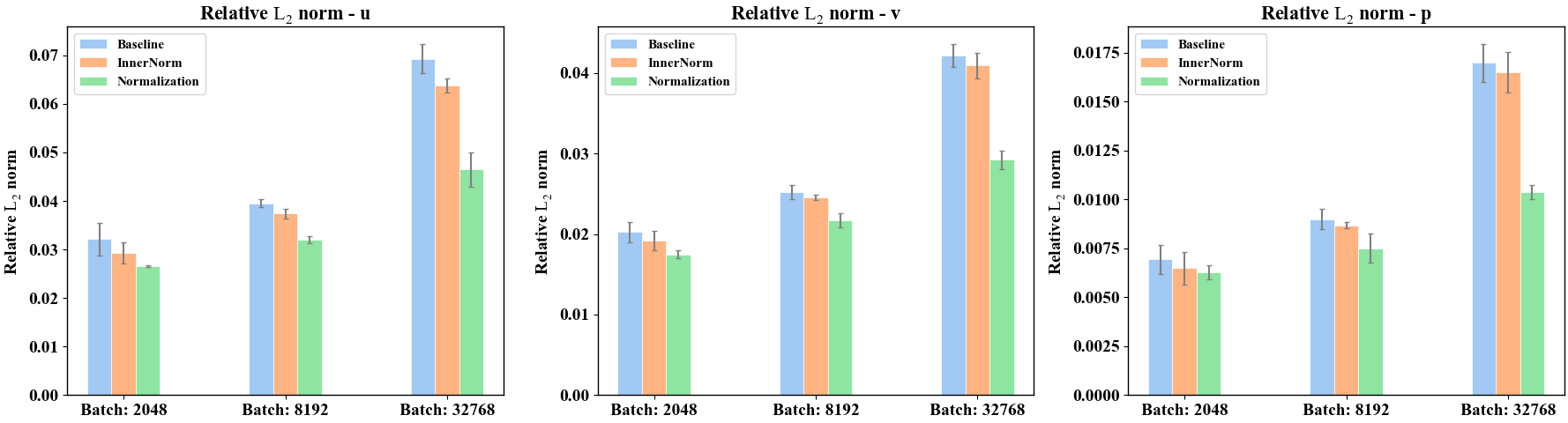}}
\caption{Performance of Baseline, InnerNorm and Normalization methods under different batch sizes for reconstructing flow field. Relative $L_2$ norms of \emph{Left}: stream-wise velocity $u$, \emph{Middle}: transverse velocity $v$, \emph{Right}: pressure $p$. Each result is averaged from three independent runs.}
  \label{fig:batch_size_compare}
\end{figure*}

Batch size refers to the number of residual points in one training batch, and the choice of batch size significantly influences both training time and accuracy. In this subsection, the performance of the Baseline, InnerNorm, and Normalization methods under batch sizes of 2048, 8192, and 32768 are compared, as shown in Fig.\ref{fig:batch_size_compare}. Other hyperparameters are kept constant for all training sessions, with 10 hidden layers, 32 neurons per layer, and an exponential decay learning rate schedule. Additionally, to eliminate the influence of randomness, each bar presented in Fig.\ref{fig:batch_size_compare} is averaged from three independent runs of training.
As batch size increases, the number of residual points in one training batch increases, leading to a decrease in the number of backward propagations per training epoch, which consequently results in a decline in prediction accuracy. Although this deterioration with increasing batch size is observed for all three methods and across all cases, the proposed Normalization method exhibits the smallest increase in relative $L_2$ norms compared to the other two methods.

The InnerNorm method shows minor improvement compared to the Baseline method, indicating that adding an inner normalization layer can alleviate the learning process to some extent. However, under certain circumstances, InnerNorm method yields worse prediction accuracy than Baseline method. As demonstrated in Fig.\ref{fig:batch_size_compare_aveRe10000}, when the batch size is 8192, the average relative $L_2$ norms of InnerNorm method is higher than that of Baseline method in predicting the stream-wise velocity $u$ and transverse velocity $v$ of cylinder wake at $\rm{Re}=10000$. In contrast, the proposed Normalization method, across all three cases, consistently exhibits lower relative $L_2$ norms in predicting all three flow quantities ${u,v,p}$, regardless of the batch size. Although the magnitude of improvement varies depending on different cases and flow quantities, the proposed Normalization method exhibits much more distinct enhancement in prediction accuracy than InnerNorm method across all cases, highlighting its superiority.
\subsection{\label{subsec: Learning rate schedule}Learning rate schedule}
\begin{figure*}[!htbp]
  \centering
  \includegraphics[width=0.50\textwidth]{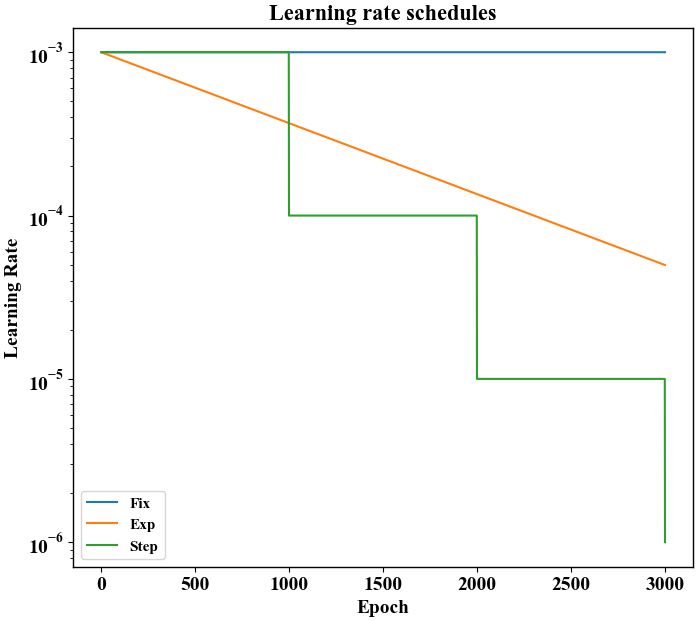}
  \caption{Learning rate schedules of fixed learning rate, exponential decay learning rate and step decay learning rate, with initial learning rate of 0.001.}
  \label{fig:lr_schedules}
\end{figure*}

\begin{figure*}[!htbp]
  \centering
\subfigure[Case 1: Cylinder wake at $\rm{Re}=3900$]{\label{fig:learning_rate_compare_aveRe3900}
\includegraphics[width=0.95\linewidth]{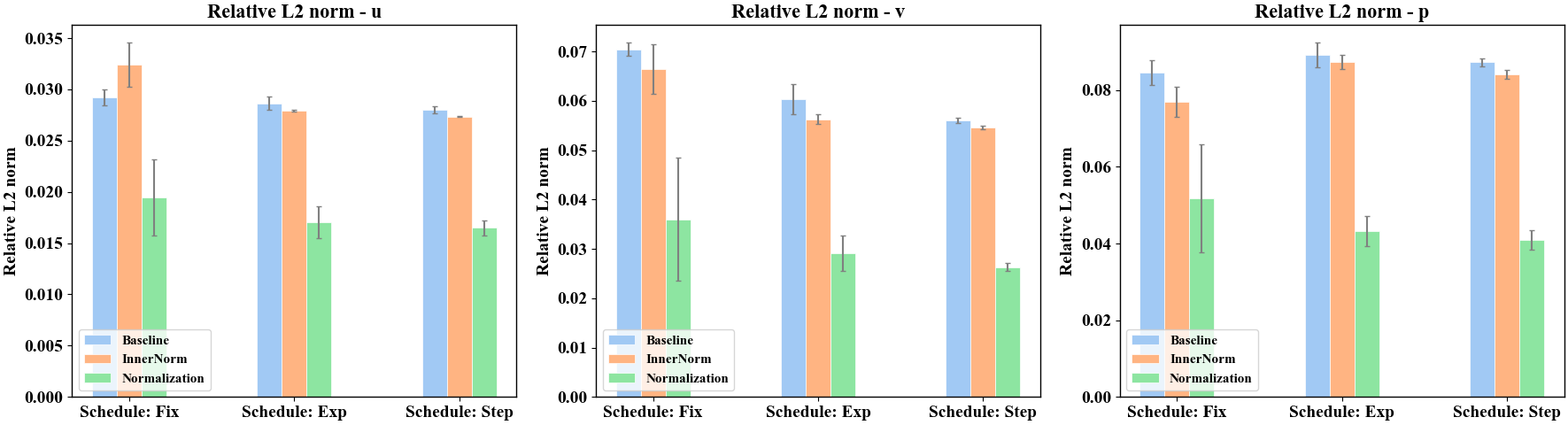}}
\hfill
\subfigure[Case 2: Cylinder wake at $\rm{Re}=10000$]{\label{fig:learning_rate_compare_aveRe10000}
\includegraphics[width=0.95\linewidth]{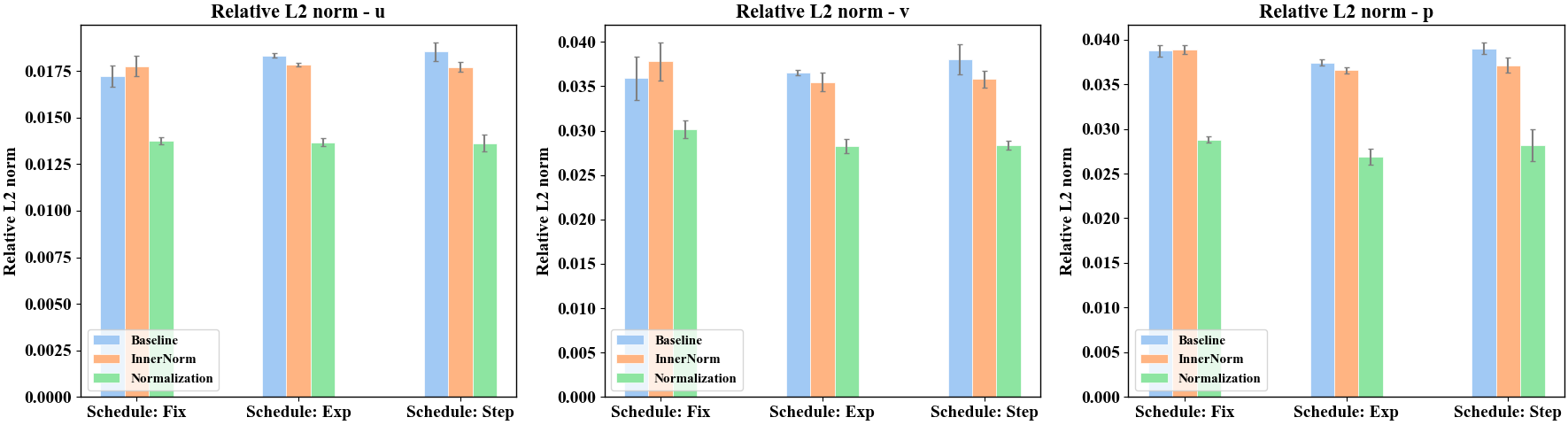}}
\hfill
\subfigure[Case 3: Decaying turbulence at $\rm{Re}=2000$]{\label{fig:learning_rate_compare_aveRe2000}
\includegraphics[width=0.95\linewidth]{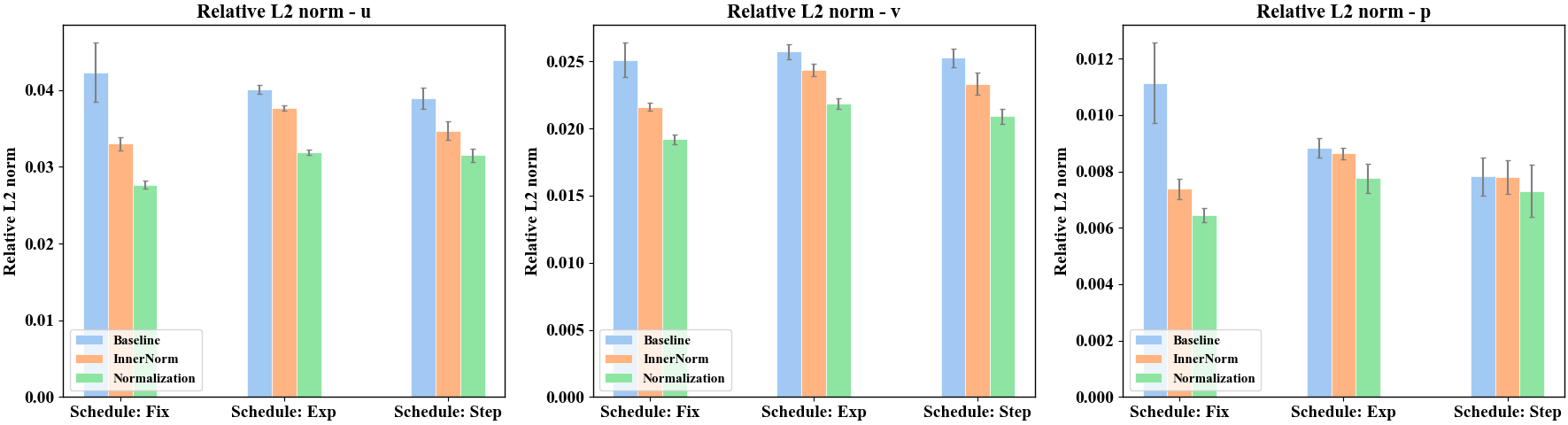}}
\caption{Performance of Baseline, InnerNorm and Normalization methods under different learning rate schedules for reconstructing flow field. Relative $L_2$ norms of \emph{Left}: stream-wise velocity $u$, \emph{Middle}: transverse velocity $v$, \emph{Right}: pressure $p$. Each result is averaged from three independent runs.}
  \label{fig:learning_rate_compare}
\end{figure*}

The learning rate schedule dictates how the learning rate adjusts throughout the training process. Commonly utilized schedules include fixed learning rate, exponential decay learning rate and step decay learning rate, as shown in Fig.\ref{fig:lr_schedules}. 
Similar to subsection \ref{subsec: Batch size}, to rule out the influence of randomness and conclusively compare the performance of the Baseline, InnerNorm, and Normalization methods under different learning rate schedules, this subsection conducts three independent training runs for all three methods across all three cases using the aforementioned learning rate schedules. For all training sessions, other hyperparameters are kept constant: the number of hidden layers is 10, number of hidden neurons in each layer is 32, batch size is 8192. The average and standard deviation values are presented in Fig.\ref{fig:learning_rate_compare}.

While the best learning rate schedule for different cases and different flow quantities varies with the choice of specific method, the superiority of the proposed Normalization method is evident: it consistently demonstrates the highest prediction accuracy for all flow quantities across all test cases.
\subsection{\label{subsec: Parameter size}Parameter size}
\begin{figure*}[!htbp]
  \centering
\subfigure[]{\label{fig:parameter_size_compare_Re3900_u}
\includegraphics[width=0.78\textwidth]{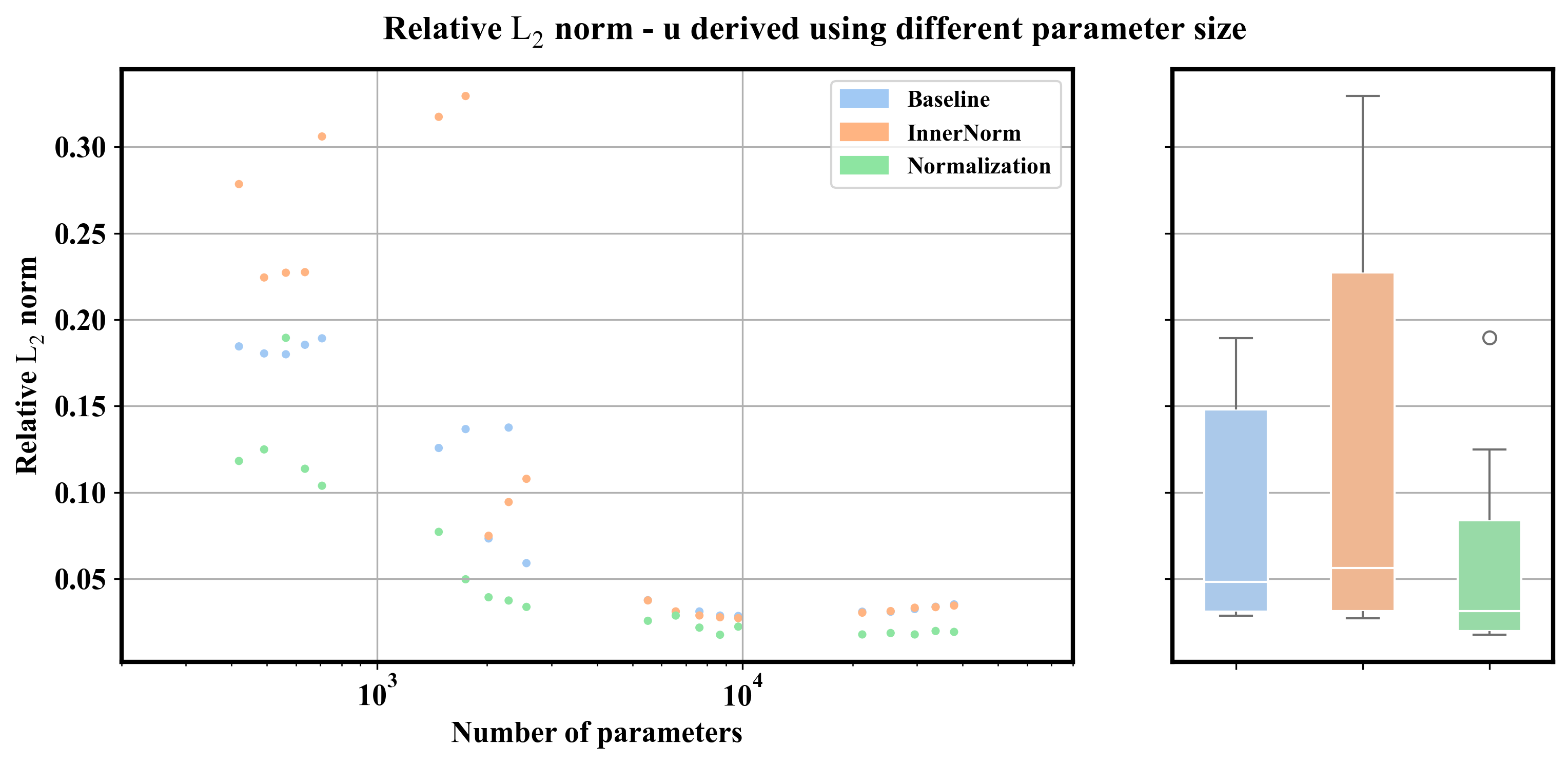}}
\hfill
\subfigure[]{\label{fig:parameter_size_compare_Re3900_v}
\includegraphics[width=0.78\textwidth]{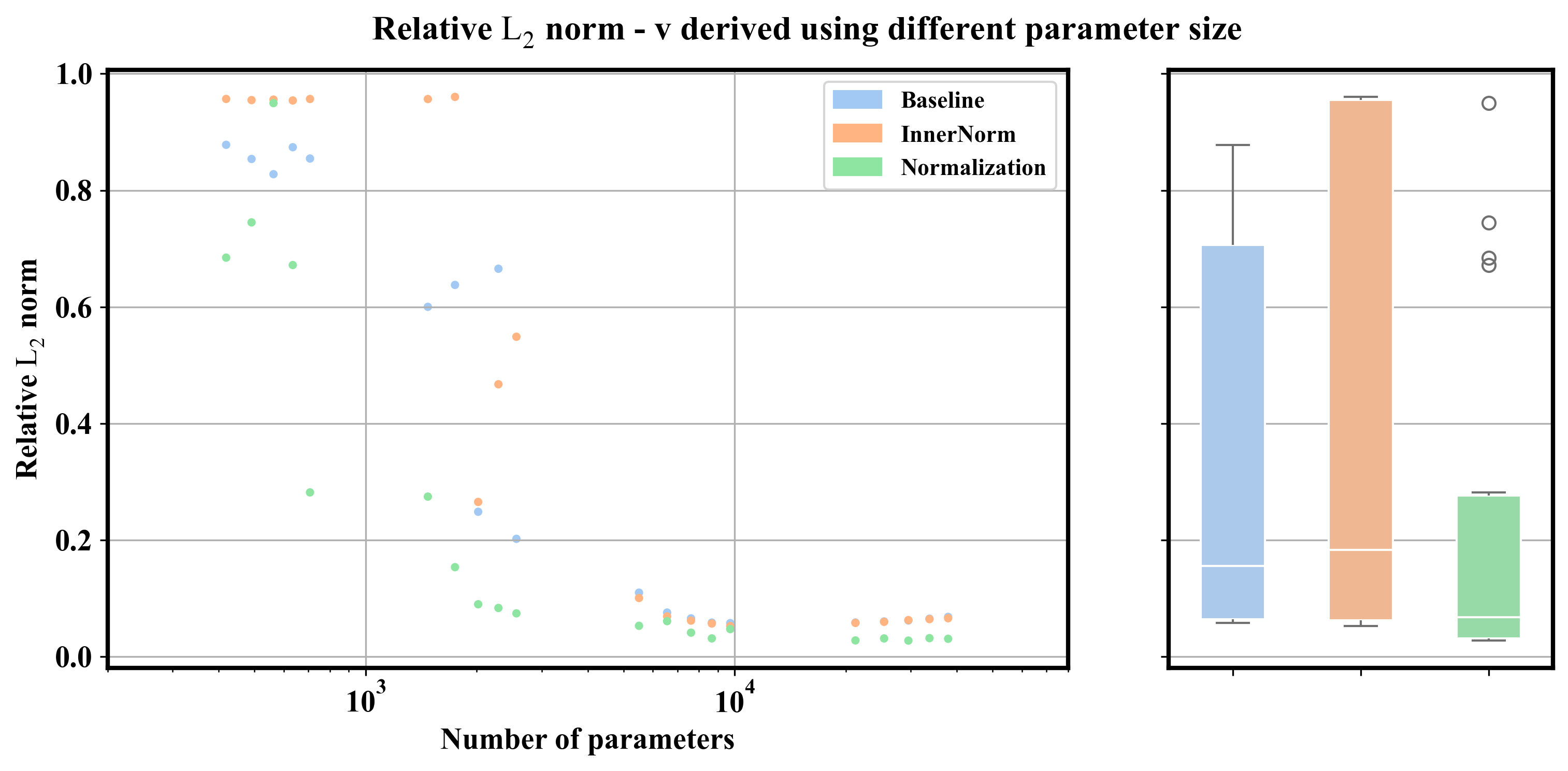}}
\hfill
\subfigure[]{\label{fig:parameter_size_compare_Re3900_p}
\includegraphics[width=0.78\textwidth]{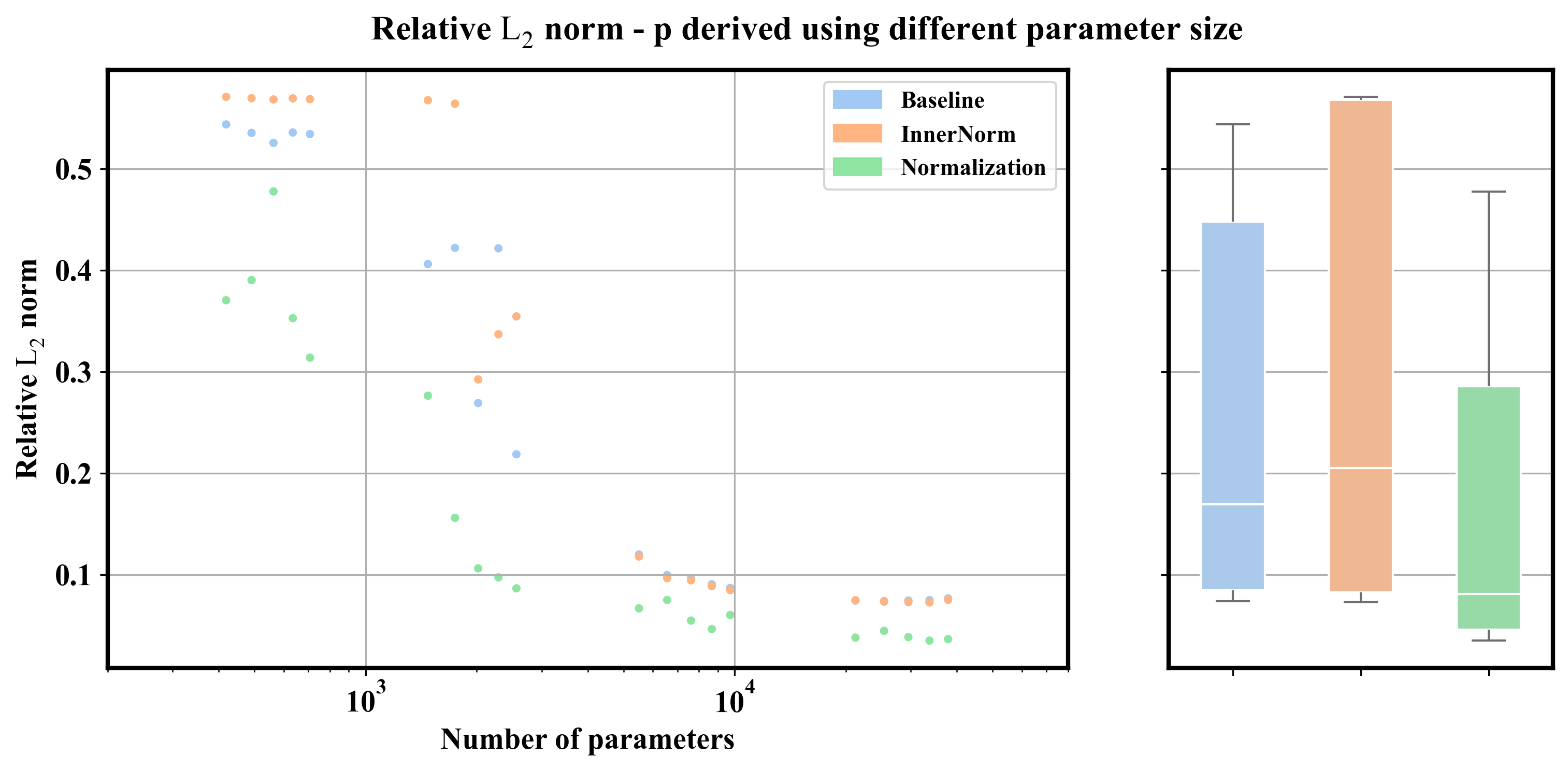}}
\caption{Performance of Baseline, InnerNorm and Normalization methods under different parameter sizes for reconstructing flow field of cylinder wake at $\rm{Re}=3900$ (Case 1). Scatter plot and box plot of relative $L_2$ norms for (a): stream-wise velocity $u$, (b): transverse velocity $v$, (c): pressure $p$.}
  \label{fig:parameter_size_compare_Re3900}
\end{figure*}

\begin{figure*}[!htbp]
  \centering
\subfigure[]{\label{fig:parameter_size_compare_Re10000_u}
\includegraphics[width=0.78\textwidth]{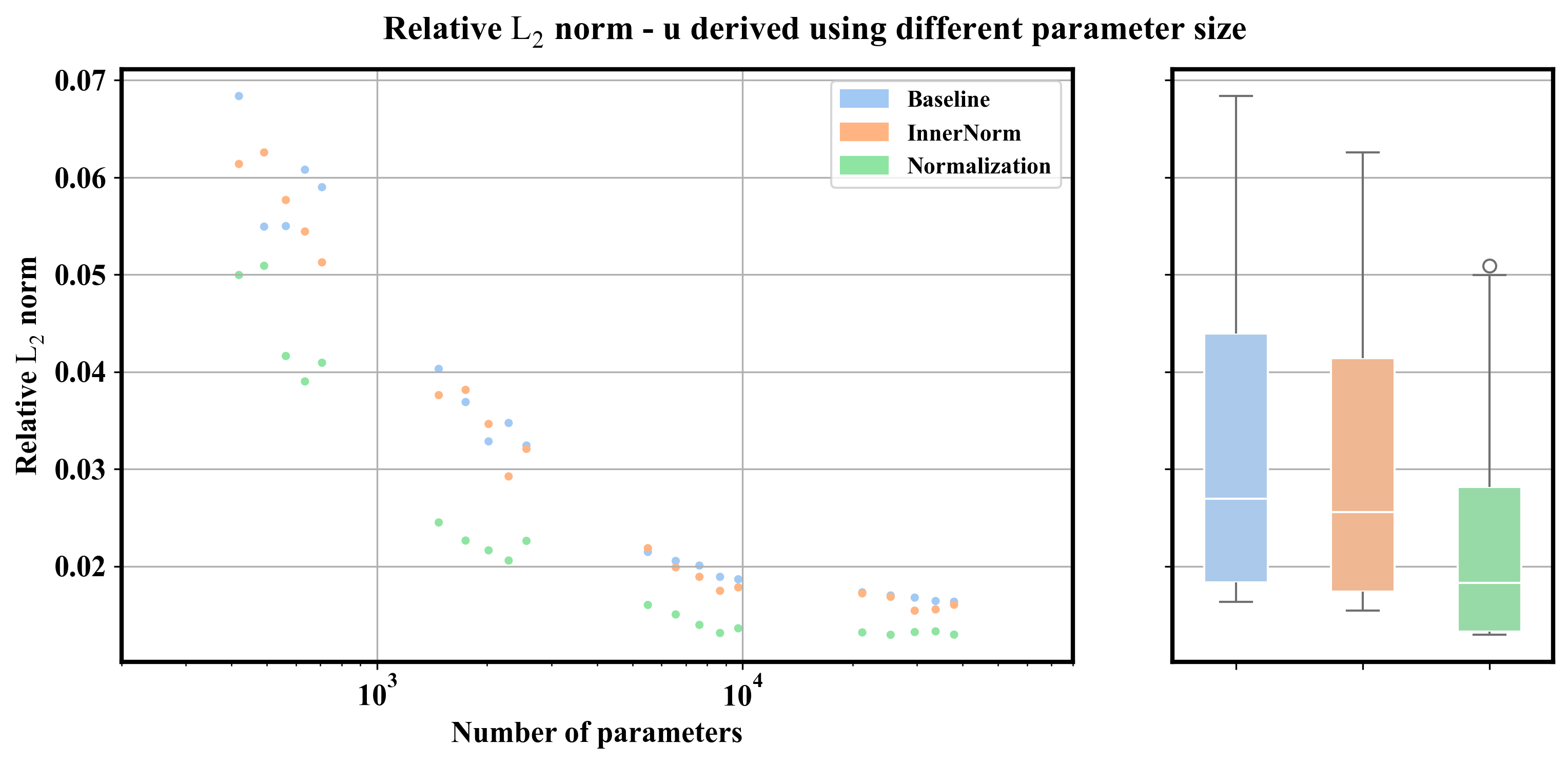}}
\hfill
\subfigure[]{\label{fig:parameter_size_compare_Re10000_v}
\includegraphics[width=0.78\textwidth]{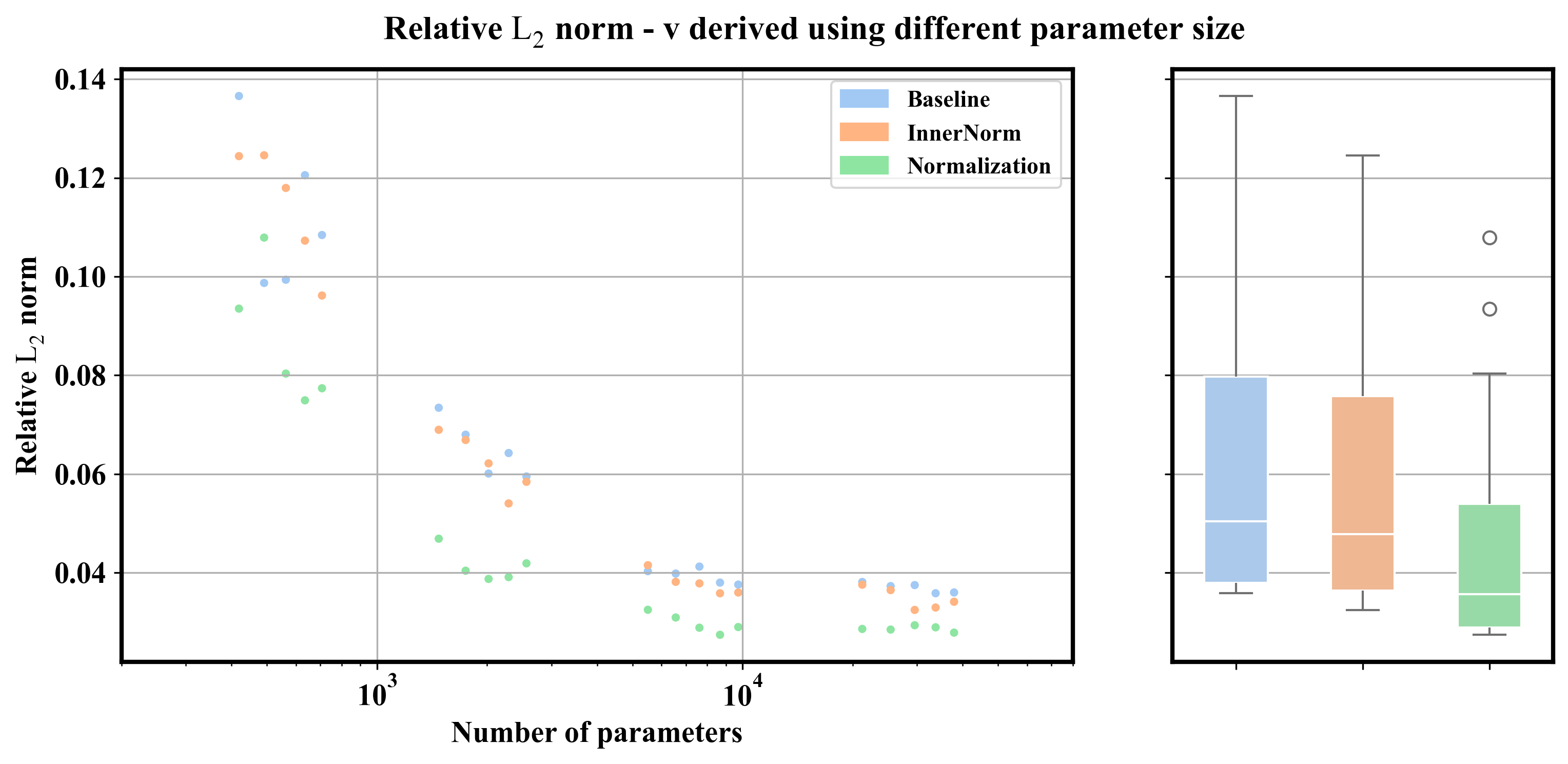}}
\hfill
\subfigure[]{\label{fig:parameter_size_compare_Re10000_p}
\includegraphics[width=0.78\textwidth]{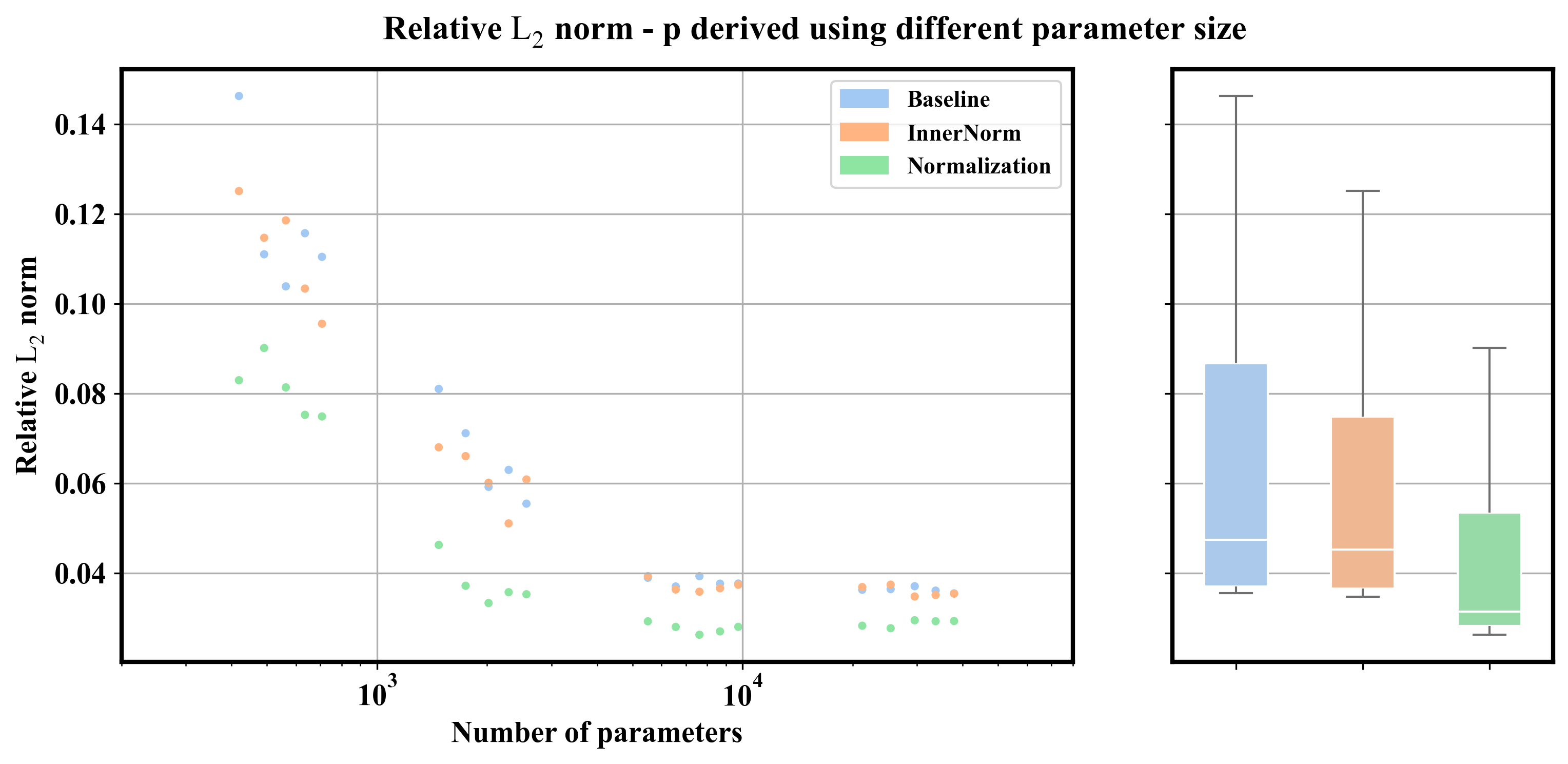}}
\caption{Performance of Baseline, InnerNorm and Normalization methods under different parameter sizes for reconstructing flow field of cylinder wake at $\rm{Re}=10000$ (Case 2). Scatter plot and box plot of relative $L_2$ norms for (a): stream-wise velocity $u$, (b): transverse velocity $v$, (c): pressure $p$.}
  \label{fig:parameter_size_compare_Re10000}
\end{figure*}

\begin{figure*}[!htbp]
  \centering
\subfigure[]{\label{fig:parameter_size_compare_Re2000_u}
\includegraphics[width=0.78\textwidth]{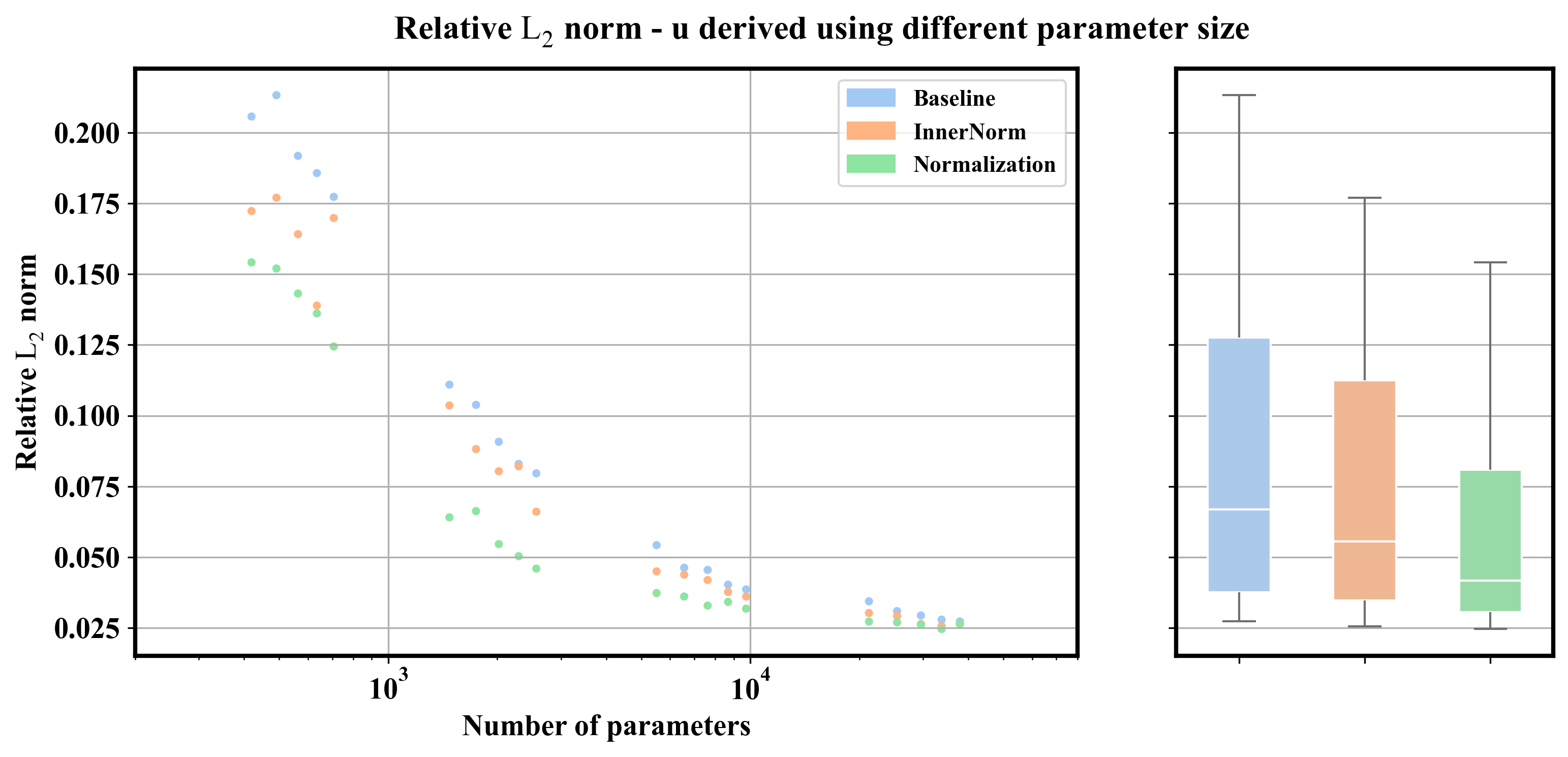}}
\hfill
\subfigure[]{\label{fig:parameter_size_compare_Re2000_v}
\includegraphics[width=0.78\textwidth]{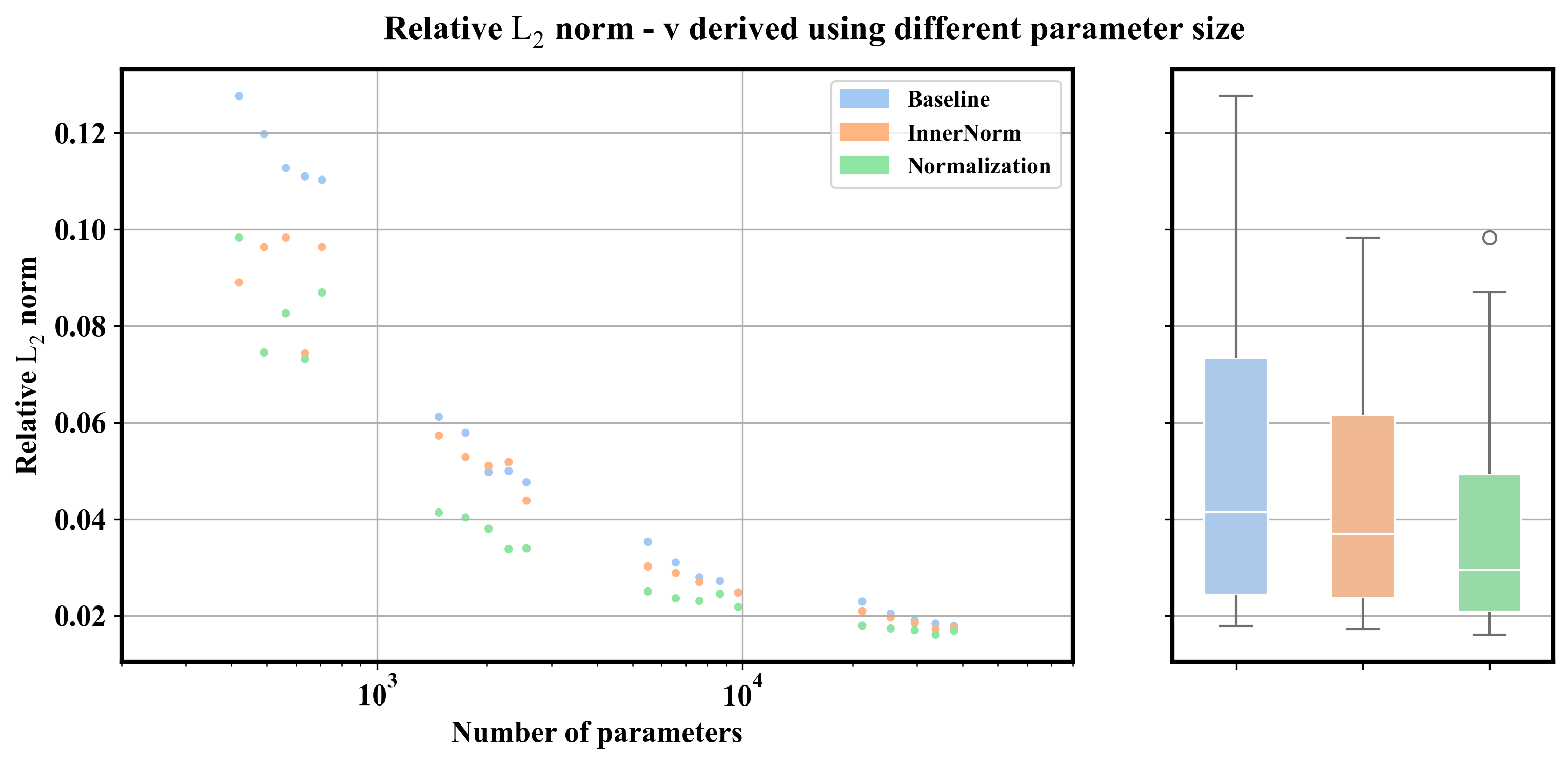}}
\hfill
\subfigure[]{\label{fig:parameter_size_compare_Re2000_p}
\includegraphics[width=0.78\textwidth]{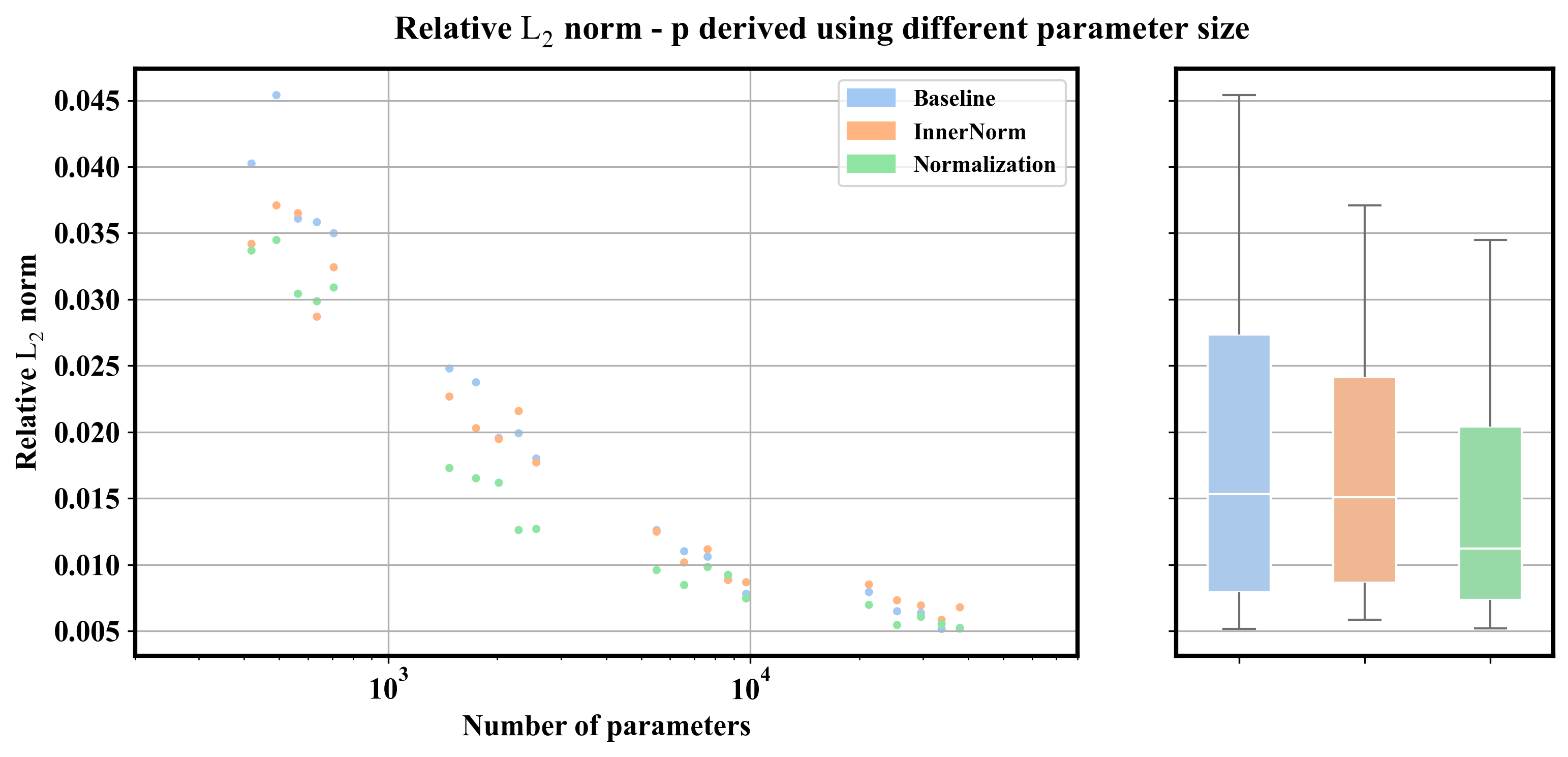}}
\caption{Performance of Baseline, InnerNorm and Normalization methods under different parameter sizes for reconstructing flow field of decaying turbulence at $\rm{Re}=2000$ (Case 3). Scatter plot and box plot of relative $L_2$ norms for (a): stream-wise velocity $u$, (b): transverse velocity $v$, (c): pressure $p$.}
  \label{fig:parameter_size_compare_Re2000}
\end{figure*}

Parameter size, defined by the number of hidden layers and neurons in each layer, represents the total number of parameters—both weights and biases—in the neural network. It is one of the most important hyperparameters to be played with when training a neural network. Generally, a larger parameter size enhances the expressiveness of the neural network. In this subsection, different numbers of hidden layers ${6,7,8,9,10}$ and hidden neurons ${8, 16, 32, 64}$ are combined to constitute different parameter size, ranging from 419 to 37891. For consistency, the learning rate schedule and batch size remain constant across all training sessions, employing an exponential decay learning rate and a batch size of 8192. Figures \ref{fig:parameter_size_compare_Re3900}, \ref{fig:parameter_size_compare_Re10000}, and \ref{fig:parameter_size_compare_Re2000} illustrate the performance of the Baseline, InnerNorm, and Normalization methods for reconstructing flow fields across different parameter sizes for all three cases.

As observed from the box plots, although InnerNorm method does present better prediction results compared to Baseline method in Case 2 and Case 3, it fails to outperform Baseline method in Case 1. In contrast, the proposed Normalization method exhibits consistently lower range and mean value of relative $L_2$ norms in all flow quantities across all three cases. Additionally, as indicated by the scatter plots, while prediction accuracy fluctuates significantly with the number of parameters, the proposed Normalization method consistently displays a lower value of relative $L_2$ norms regardless of parameter size. Moreover, when the choice of parameter size is not appropriate (the number of parameters is extremely small, only several hundred), the advantage of the proposed Normalization method is even more pronounced compared to the other two methods, suggesting Normalization can always be a "better and safer choice" when exploring the hyperparameter space.
\section{\label{sec: ANALYSIS}ANALYSIS}
\begin{figure*}[!htbp]
  \centering
  \includegraphics[width=1.00\textwidth]{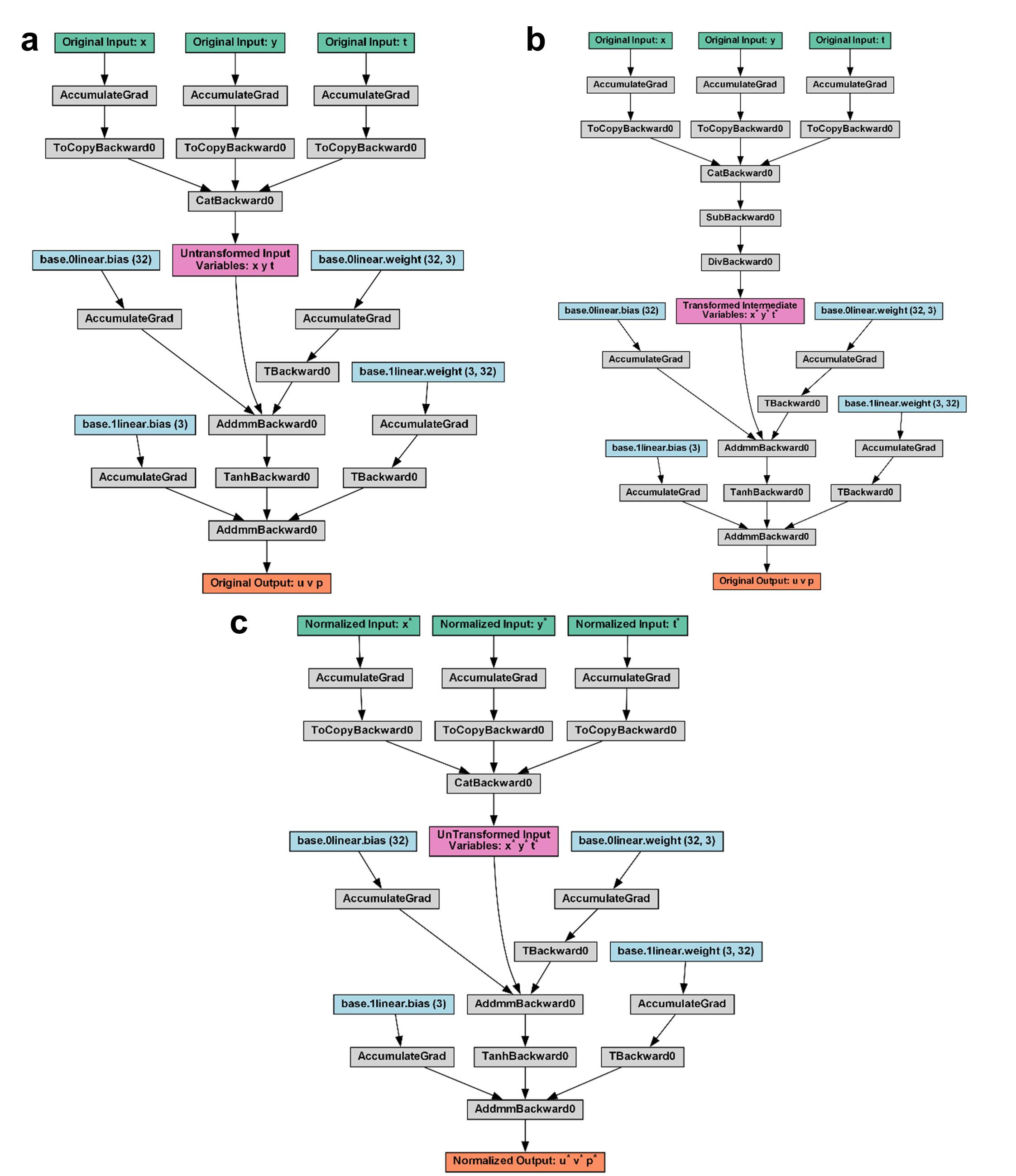}
  \caption{Computational graph of Baseline, InnerNorm, and Normalization method, taking neural network with a single hidden layer and 32 hidden neurons as an example. The green boxes represent network inputs, orange boxes indicate network outputs, and pink boxes denote transformed or untransformed intermediate variables. The blue boxes signify weights and biases that evolve with the training process. Since the gray boxes represent fixed forward and backward computations that remain unchanged by training, the actual learned mapping relationship is established between the pink and orange boxes.}
  \label{fig:computational_graph}
\end{figure*}

An intriguing question needs to be addressed is why the Normalization method outperforms both the Baseline and InnerNorm methods. However, it is hard to give an explicit quantitative analysis, especially considering the black-box nature of neural networks. In this section, we present a potential analysis from the perspective of computational graph. For illustration, the computational graphs for all three methods using a one-hidden-layer network are depicted in Fig.\ref{fig:computational_graph}.

Although, at first glance, it appears that the mapping relationship learned by PINNs is established between network inputs and network outputs, the actual mapping relationship is established between transformed or untransformed intermediate variables (pink boxes in Fig.\ref{fig:computational_graph}) and network outputs. This is because the gray boxes represent fixed forward and backward computations that remain unchanged by network training. Therefore, the actual mapping relationships learned by the Baseline, InnerNorm, and Normalization methods are $\{ x,y,t\} \to \{ u,v,p\}$, $\{ {x^*},{y^*},{t^*}\} \to \{ u,v,p\}$, and $\{ {x^*},{y^*},{t^*}\} \to \{ {u^*},{v^*},{p^*}\}$, respectively. As discussed in subsection \ref{subsec:normalization}, more concentrated features contribute to better learning performance. This partially explains the unstable yet prevailing advantage of the InnerNorm method over the Baseline method, as the features of spatiotemporal coordinates are more concentrated. However, while adding an inner normalization layer is straightforward, it is not as intuitive as the proposed Normalization method, which constructs the mapping relationship directly based on both normalized inputs and outputs. From this perspective, the superiority of the proposed Normalization method compared to the Baseline and InnerNorm methods is reasonable and predictable, as normalizing all the features provides more concentrated scales.
\section{\label{sec: CONCLUSION AND DISCUSSION}CONCLUSION AND DISCUSSION}
To summarize, by taking reference from classic machine learning methods and making analogy to non-dimensionalization, we implemented the normalization of NS equations and proposed a data preprocessing method for physics-informed neural networks, with a specific focus on data normalization. Unlike the existing approach of adding an inner normalization layer, our normalization method involves directly normalizing both input and output data based on statistical information of the training dataset. The neural networks are then constructed using these normalized features, aligning with the prevalent data normalization techniques in classic machine learning. By concurrently normalizing the data and transforming the corresponding equations, the inverse problem of flow field reconstruction was better solved than simply applying non-dimensional equations or adding an extra inner normalization layer. In the context of three turbulent flow cases - the wake flow past 2D circular cylinder at $\rm{Re}=3900$, the wake flow past 2D circular cylinder at $\rm{Re}=10000$, and 2D decaying turbulence at $\rm{Re}=2000$, the proposed normalization method consistently yield lower relative $L_2$ norm values for each flow variable $u,v,p$ under different batch sizes, learning rate schedules, and parameter sizes. This improvement was achieved despite the variations in Reynolds numbers and turbulence models used in the simulations.

While the inner normalization layer method exhibited some enhancement in prediction accuracy for three test cases, the gains were marginal and often negligible, with some cases demonstrating worsened performance when compared to the baseline method of solely applying non-dimensional equations. Conversely, the intuitive and mathematically-based normalization method exhibited much more prominent advantage in predicting all flow quantities, highlighting its validity and superiority. Despite utilizing only 36 sparsely distributed data points, the mean and standard deviation values derived from this limited dataset effectively showcased the method's ability to optimize data utilization, hinting at its potential applicability for larger datasets in inverse problems.

Although the present study solely validated the normalization method on the 2D unsteady NS equations, the underlying mathematical principles of the calculus technique employed to transform these equations are universally applicable to all types of differential equations. Whether dealing with dimensional or non-dimensional equations, ordinary or partial differential equations, they can all undergo normalization using a consistent preprocessing approach. We actively anticipate widespread adoption of the proposed normalization method across diverse types of differential equations, potentially establishing itself as an essential preprocessing technique for training PINNs.
\section*{ACKNOWLEDGMENTS}
This work was supported by National Key Research and Development Project under 2022YFB2603400, China National Railway Group Science and Technology Program (Grant No. K2023J047), and the International Partnership Program of Chinese Academy of Sciences (Grant No. 025GJHZ2022118FN).
\section*{DATA AVAILABILITY}
The data and code that support the findings of this study will soon be available on \url{https://github.com/Shengfeng233/PINN-Preprocess}.
\newpage
\bibliographystyle{unsrt}  

\bibliography{references}

\providecommand{\noopsort}[1]{}\providecommand{\singleletter}[1]{#1}%
\begin{thebibliography}{10}

\bibitem{kiran2021deep}
B~Ravi Kiran, Ibrahim Sobh, Victor Talpaert, Patrick Mannion, Ahmad~A
  Al~Sallab, Senthil Yogamani, and Patrick P{\'e}rez.
\newblock Deep reinforcement learning for autonomous driving: A survey.
\newblock {\em IEEE Transactions on Intelligent Transportation Systems},
  23(6):4909--4926, 2021.

\bibitem{portugal2018use}
Ivens Portugal, Paulo Alencar, and Donald Cowan.
\newblock The use of machine learning algorithms in recommender systems: A
  systematic review.
\newblock {\em Expert Systems with Applications}, 97:205--227, 2018.

\bibitem{morgan2020opportunities}
Dane Morgan and Ryan Jacobs.
\newblock Opportunities and challenges for machine learning in materials
  science.
\newblock {\em Annual Review of Materials Research}, 50:71--103, 2020.

\bibitem{greener2022guide}
Joe~G Greener, Shaun~M Kandathil, Lewis Moffat, and David~T Jones.
\newblock A guide to machine learning for biologists.
\newblock {\em Nature Reviews Molecular Cell Biology}, 23(1):40--55, 2022.

\bibitem{brunton2020machine}
Steven~L Brunton, Bernd~R Noack, and Petros Koumoutsakos.
\newblock Machine learning for fluid mechanics.
\newblock {\em Annual review of fluid mechanics}, 52:477--508, 2020.

\bibitem{karniadakis2021physics}
George~Em Karniadakis, Ioannis~G Kevrekidis, Lu~Lu, Paris Perdikaris, Sifan
  Wang, and Liu Yang.
\newblock Physics-informed machine learning.
\newblock {\em Nature Reviews Physics}, 3(6):422--440, 2021.

\bibitem{cuomo2022scientific}
Salvatore Cuomo, Vincenzo~Schiano Di~Cola, Fabio Giampaolo, Gianluigi Rozza,
  Maziar Raissi, and Francesco Piccialli.
\newblock Scientific machine learning through physics--informed neural
  networks: where we are and what’s next.
\newblock {\em Journal of Scientific Computing}, 92(3):88, 2022.

\bibitem{lagaris1998artificial}
Isaac~E Lagaris, Aristidis Likas, and Dimitrios~I Fotiadis.
\newblock Artificial neural networks for solving ordinary and partial
  differential equations.
\newblock {\em IEEE transactions on neural networks}, 9(5):987--1000, 1998.

\bibitem{raissi2019physics}
Maziar Raissi, Paris Perdikaris, and George~E Karniadakis.
\newblock Physics-informed neural networks: A deep learning framework for
  solving forward and inverse problems involving nonlinear partial differential
  equations.
\newblock {\em Journal of Computational physics}, 378:686--707, 2019.

\bibitem{yang2021b}
Liu Yang, Xuhui Meng, and George~Em Karniadakis.
\newblock B-pinns: Bayesian physics-informed neural networks for forward and
  inverse pde problems with noisy data.
\newblock {\em Journal of Computational Physics}, 425:109913, 2021.

\bibitem{yu2022gradient}
Jeremy Yu, Lu~Lu, Xuhui Meng, and George~Em Karniadakis.
\newblock Gradient-enhanced physics-informed neural networks for forward and
  inverse pde problems.
\newblock {\em Computer Methods in Applied Mechanics and Engineering},
  393:114823, 2022.

\bibitem{zhang2023enforcing}
Zhi-Yong Zhang, Hui Zhang, Li-Sheng Zhang, and Lei-Lei Guo.
\newblock Enforcing continuous symmetries in physics-informed neural network
  for solving forward and inverse problems of partial differential equations.
\newblock {\em Journal of Computational Physics}, 492:112415, 2023.

\bibitem{cai2021physics}
Shengze Cai, Zhicheng Wang, Sifan Wang, Paris Perdikaris, and George~Em
  Karniadakis.
\newblock Physics-informed neural networks for heat transfer problems.
\newblock {\em Journal of Heat Transfer}, 143(6), 2021.

\bibitem{jagtap2023coolpinns}
Nimish~V Jagtap, MK~Mudunuru, and KB~Nakshatrala.
\newblock Coolpinns: A physics-informed neural network modeling of active
  cooling in vascular systems.
\newblock {\em Applied Mathematical Modelling}, 122:265--287, 2023.

\bibitem{ren2024seismicnet}
Pu~Ren, Chengping Rao, Su~Chen, Jian-Xun Wang, Hao Sun, and Yang Liu.
\newblock Seismicnet: Physics-informed neural networks for seismic wave
  modeling in semi-infinite domain.
\newblock {\em Computer Physics Communications}, 295:109010, 2024.

\bibitem{zhou2021solving}
Zijian Zhou and Zhenya Yan.
\newblock Solving forward and inverse problems of the logarithmic nonlinear
  schr{\"o}dinger equation with pt-symmetric harmonic potential via deep
  learning.
\newblock {\em Physics Letters A}, 387:127010, 2021.

\bibitem{jin2022physics}
Henry Jin, Marios Mattheakis, and Pavlos Protopapas.
\newblock Physics-informed neural networks for quantum eigenvalue problems.
\newblock In {\em 2022 International Joint Conference on Neural Networks
  (IJCNN)}, pages 1--8. IEEE, 2022.

\bibitem{cai2021fluid}
Shengze Cai, Zhiping Mao, Zhicheng Wang, Minglang Yin, and George~Em
  Karniadakis.
\newblock Physics-informed neural networks (pinns) for fluid mechanics: A
  review.
\newblock {\em Acta Mechanica Sinica}, 37(12):1727--1738, 2021.

\bibitem{sharma2023review}
Pushan Sharma, Wai~Tong Chung, Bassem Akoush, and Matthias Ihme.
\newblock A review of physics-informed machine learning in fluid mechanics.
\newblock {\em Energies}, 16(5):2343, 2023.

\bibitem{rao2020physics}
Chengping Rao, Hao Sun, and Yang Liu.
\newblock Physics-informed deep learning for incompressible laminar flows.
\newblock {\em Theoretical and Applied Mechanics Letters}, 10(3):207--212,
  2020.

\bibitem{biswas2023three}
Saykat~Kumar Biswas and NK~Anand.
\newblock Three-dimensional laminar flow using physics informed deep neural
  networks.
\newblock {\em Physics of Fluids}, 35(12), 2023.

\bibitem{jin2021nsfnets}
Xiaowei Jin, Shengze Cai, Hui Li, and George~Em Karniadakis.
\newblock Nsfnets (navier-stokes flow nets): Physics-informed neural networks
  for the incompressible navier-stokes equations.
\newblock {\em Journal of Computational Physics}, 426:109951, 2021.

\bibitem{hanrahan2023studying}
S~Hanrahan, M~Kozul, and RD~Sandberg.
\newblock Studying turbulent flows with physics-informed neural networks and
  sparse data.
\newblock {\em International Journal of Heat and Fluid Flow}, 104:109232, 2023.

\bibitem{raissi2020hidden}
Maziar Raissi, Alireza Yazdani, and George~Em Karniadakis.
\newblock Hidden fluid mechanics: Learning velocity and pressure fields from
  flow visualizations.
\newblock {\em Science}, 367(6481):1026--1030, 2020.

\bibitem{cai2021flow}
Shengze Cai, Zhicheng Wang, Frederik Fuest, Young~Jin Jeon, Callum Gray, and
  George~Em Karniadakis.
\newblock Flow over an espresso cup: inferring 3-d velocity and pressure fields
  from tomographic background oriented schlieren via physics-informed neural
  networks.
\newblock {\em Journal of Fluid Mechanics}, 915:A102, 2021.

\bibitem{fathi2020super}
Mojtaba~F Fathi, Isaac Perez-Raya, Ahmadreza Baghaie, Philipp Berg, Gabor
  Janiga, Amirhossein Arzani, and Roshan~M D’Souza.
\newblock Super-resolution and denoising of 4d-flow mri using physics-informed
  deep neural nets.
\newblock {\em Computer Methods and Programs in Biomedicine}, 197:105729, 2020.

\bibitem{baydin2018automatic}
Atilim~Gunes Baydin, Barak~A Pearlmutter, Alexey~Andreyevich Radul, and
  Jeffrey~Mark Siskind.
\newblock Automatic differentiation in machine learning: a survey.
\newblock {\em Journal of Marchine Learning Research}, 18:1--43, 2018.

\bibitem{sun2020surrogate}
Luning Sun, Han Gao, Shaowu Pan, and Jian-Xun Wang.
\newblock Surrogate modeling for fluid flows based on physics-constrained deep
  learning without simulation data.
\newblock {\em Computer Methods in Applied Mechanics and Engineering},
  361:112732, 2020.

\bibitem{xu2023practical}
Shengfeng Xu, Zhenxu Sun, Renfang Huang, Dilong Guo, Guowei Yang, and Shengjun
  Ju.
\newblock A practical approach to flow field reconstruction with sparse or
  incomplete data through physics informed neural network.
\newblock {\em Acta Mechanica Sinica}, 39(3):1--15, 2023.

\bibitem{xu2021explore}
Hui Xu, Wei Zhang, and Yong Wang.
\newblock Explore missing flow dynamics by physics-informed deep learning: The
  parameterized governing systems.
\newblock {\em Physics of Fluids}, 33(9):095116, 2021.

\bibitem{wang2024dynamic}
Longyan Wang, Meng Chen, Zhaohui Luo, Bowen Zhang, Jian Xu, Zilu Wang, and
  Andy~CC Tan.
\newblock Dynamic wake field reconstruction of wind turbine through
  physics-informed neural network and sparse lidar data.
\newblock {\em Energy}, page 130401, 2024.

\bibitem{xu2023spatiotemporal}
Shengfeng Xu, Chang Yan, Guangtao Zhang, Zhenxu Sun, Renfang Huang, Shengjun
  Ju, Dilong Guo, and Guowei Yang.
\newblock Spatiotemporal parallel physics-informed neural networks: A framework
  to solve inverse problems in fluid mechanics.
\newblock {\em Physics of Fluids}, 35(6), 2023.

\bibitem{liu2024high}
Shiyu Liu, Haiou Wang, Jacqueline~H Chen, Kun Luo, and Jianren Fan.
\newblock High-resolution reconstruction of turbulent flames from sparse data
  with physics-informed neural networks.
\newblock {\em Combustion and Flame}, 260:113275, 2024.

\bibitem{jagtap2020conservative}
Ameya~D Jagtap, Ehsan Kharazmi, and George~Em Karniadakis.
\newblock Conservative physics-informed neural networks on discrete domains for
  conservation laws: Applications to forward and inverse problems.
\newblock {\em Computer Methods in Applied Mechanics and Engineering},
  365:113028, 2020.

\bibitem{hu2023augmented}
Zheyuan Hu, Ameya~D Jagtap, George~Em Karniadakis, and Kenji Kawaguchi.
\newblock Augmented physics-informed neural networks (apinns): A gating
  network-based soft domain decomposition methodology.
\newblock {\em Engineering Applications of Artificial Intelligence},
  126:107183, 2023.

\bibitem{meng2020ppinn}
Xuhui Meng, Zhen Li, Dongkun Zhang, and George~Em Karniadakis.
\newblock Ppinn: Parareal physics-informed neural network for time-dependent
  pdes.
\newblock {\em Computer Methods in Applied Mechanics and Engineering},
  370:113250, 2020.

\bibitem{wang2024respecting}
Sifan Wang, Shyam Sankaran, and Paris Perdikaris.
\newblock Respecting causality for training physics-informed neural networks.
\newblock {\em Computer Methods in Applied Mechanics and Engineering},
  421:116813, 2024.

\bibitem{chiu2022can}
Pao-Hsiung Chiu, Jian~Cheng Wong, Chinchun Ooi, My~Ha Dao, and Yew-Soon Ong.
\newblock Can-pinn: A fast physics-informed neural network based on
  coupled-automatic--numerical differentiation method.
\newblock {\em Computer Methods in Applied Mechanics and Engineering},
  395:114909, 2022.

\bibitem{costabal2024delta}
Francisco~Sahli Costabal, Simone Pezzuto, and Paris Perdikaris.
\newblock $\delta$-pinns: physics-informed neural networks on complex
  geometries.
\newblock {\em Engineering Applications of Artificial Intelligence},
  127:107324, 2024.

\bibitem{jagtap2020locally}
Ameya~D Jagtap, Kenji Kawaguchi, and George Em~Karniadakis.
\newblock Locally adaptive activation functions with slope recovery for deep
  and physics-informed neural networks.
\newblock {\em Proceedings of the Royal Society A}, 476(2239):20200334, 2020.

\bibitem{jagtap2020adaptive}
Ameya~D Jagtap, Kenji Kawaguchi, and George~Em Karniadakis.
\newblock Adaptive activation functions accelerate convergence in deep and
  physics-informed neural networks.
\newblock {\em Journal of Computational Physics}, 404:109136, 2020.

\bibitem{mcclenny2023self}
Levi~D McClenny and Ulisses~M Braga-Neto.
\newblock Self-adaptive physics-informed neural networks.
\newblock {\em Journal of Computational Physics}, 474:111722, 2023.

\bibitem{xu2023transfer}
Chen Xu, Ba~Trung Cao, Yong Yuan, and G{\"u}nther Meschke.
\newblock Transfer learning based physics-informed neural networks for solving
  inverse problems in engineering structures under different loading scenarios.
\newblock {\em Computer Methods in Applied Mechanics and Engineering},
  405:115852, 2023.

\bibitem{garcia2015data}
Salvador Garc{\'\i}a, Juli{\'a}n Luengo, and Francisco Herrera.
\newblock {\em Data preprocessing in data mining}, volume~72.
\newblock Springer, 2015.

\bibitem{kotsiantis2006data}
Sotiris~B Kotsiantis, Dimitris Kanellopoulos, and Panagiotis~E Pintelas.
\newblock Data preprocessing for supervised leaning.
\newblock {\em International journal of computer science}, 1(2):111--117, 2006.

\bibitem{garcia2016big}
Salvador Garc{\'\i}a, Sergio Ram{\'\i}rez-Gallego, Juli{\'a}n Luengo,
  Jos{\'e}~Manuel Ben{\'\i}tez, and Francisco Herrera.
\newblock Big data preprocessing: methods and prospects.
\newblock {\em Big Data Analytics}, 1(1):1--22, 2016.

\bibitem{de2024physics}
Mario De~Florio, Enrico Schiassi, Francesco Calabr{\`o}, and Roberto Furfaro.
\newblock Physics-informed neural networks for 2nd order odes with sharp
  gradients.
\newblock {\em Journal of Computational and Applied Mathematics}, 436:115396,
  2024.

\bibitem{mishra2023estimates}
Siddhartha Mishra and Roberto Molinaro.
\newblock Estimates on the generalization error of physics-informed neural
  networks for approximating pdes.
\newblock {\em IMA Journal of Numerical Analysis}, 43(1):1--43, 2023.

\bibitem{maddu2022inverse}
Suryanarayana Maddu, Dominik Sturm, Christian~L M{\"u}ller, and Ivo~F
  Sbalzarini.
\newblock Inverse dirichlet weighting enables reliable training of physics
  informed neural networks.
\newblock {\em Machine Learning: Science and Technology}, 3(1):015026, 2022.

\bibitem{patankar2018numerical}
Suhas Patankar.
\newblock {\em Numerical heat transfer and fluid flow}.
\newblock CRC press, 2018.

\bibitem{yan2023exploring}
Chang Yan, Shengfeng Xu, Zhenxu Sun, Dilong Guo, Shengjun Ju, Renfang Huang,
  and Guowei Yang.
\newblock Exploring hidden flow structures from sparse data through
  deep-learning-strengthened proper orthogonal decomposition.
\newblock {\em Physics of Fluids}, 35(3):037119, 2023.

\bibitem{Lauber:2D-Turbulence-Python}
Marin Lauber.
\newblock 2d-turbulence-python, 2021.
\newblock \url{https://github.com/marinlauber/2D-Turbulence-Python}, Accessed:
  February 15, 2023.

\bibitem{hao2023pinnacle}
Zhongkai Hao, Jiachen Yao, Chang Su, Hang Su, Ziao Wang, Fanzhi Lu, Zeyu Xia,
  Yichi Zhang, Songming Liu, Lu~Lu, et~al.
\newblock Pinnacle: A comprehensive benchmark of physics-informed neural
  networks for solving pdes.
\newblock {\em arXiv preprint arXiv:2306.08827}, 2023.

\bibitem{kingma2014adam}
Diederik~P Kingma and Jimmy Ba.
\newblock Adam: A method for stochastic optimization.
\newblock {\em arXiv preprint arXiv:1412.6980}, 2014.

\end{thebibliography}
\newpage
\appendix
\titleformat{\section}[block]{\normalfont\large\bfseries}{Appendix \thesection:}{1em}{}
\section{\label{Appendix:A}Detailed Qualification}
\renewcommand{\thefigure}{A.\arabic{figure}} 
\setcounter{figure}{0}  
\begin{figure*}[!htbp]
  \centering
  \includegraphics[width=0.95\textwidth]{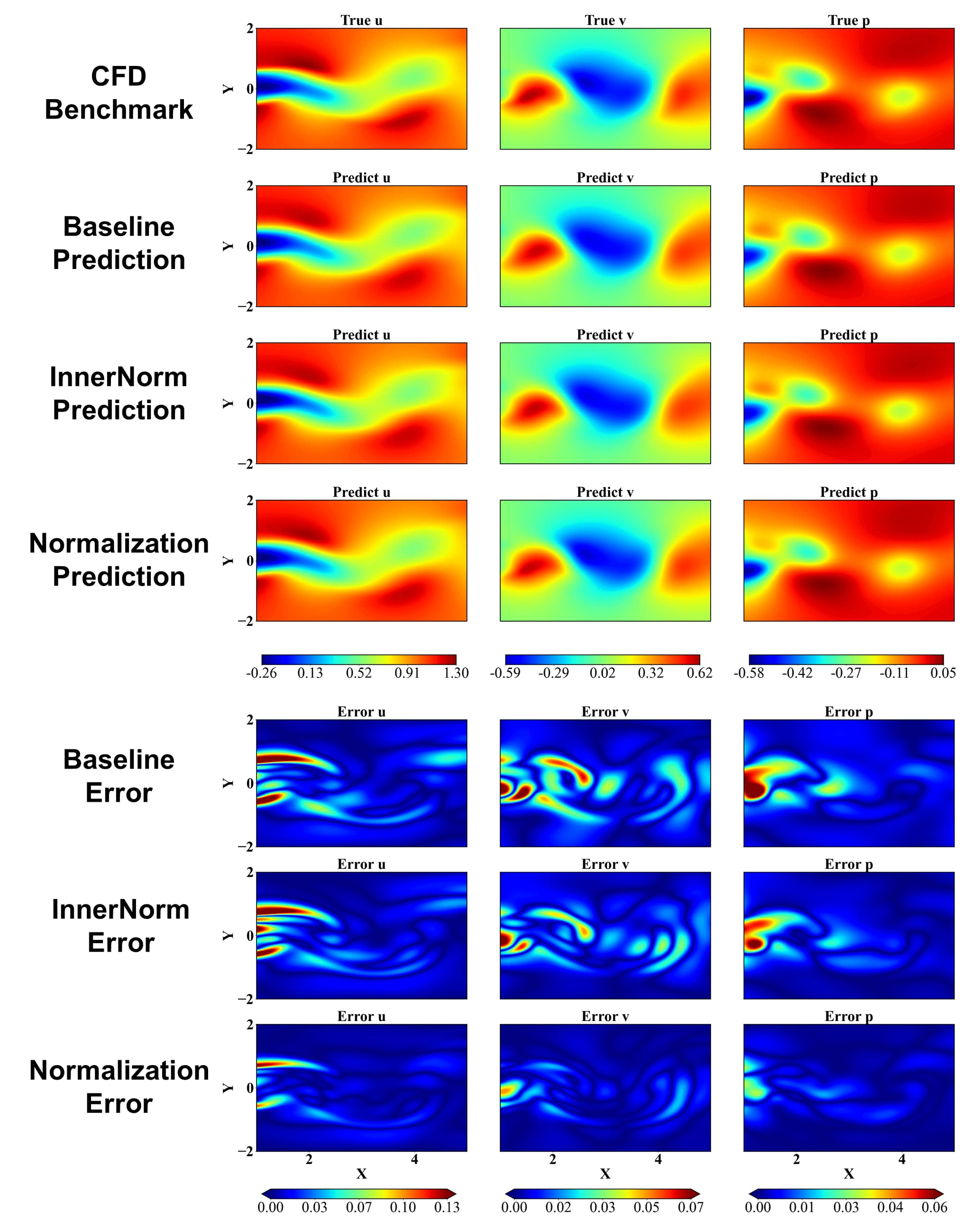}
  \caption{Flow field reconstruction result of flow past 2D circular cylinder at $\rm{Re}=3900$, at the time snapshot $t=38.17$. Contour plot and point-wise error of stream-wise velocity $u$ (first column), transverse velocity $v$ (second column) and pressure $p$ (third column). All trained with batch size of 8192, parameter size of 9731, and exponential decay learning rate.}
  \label{fig:appendixA_Re3900}
\end{figure*}
\begin{figure*}[!htbp]
  \centering
  \includegraphics[width=0.95\textwidth]{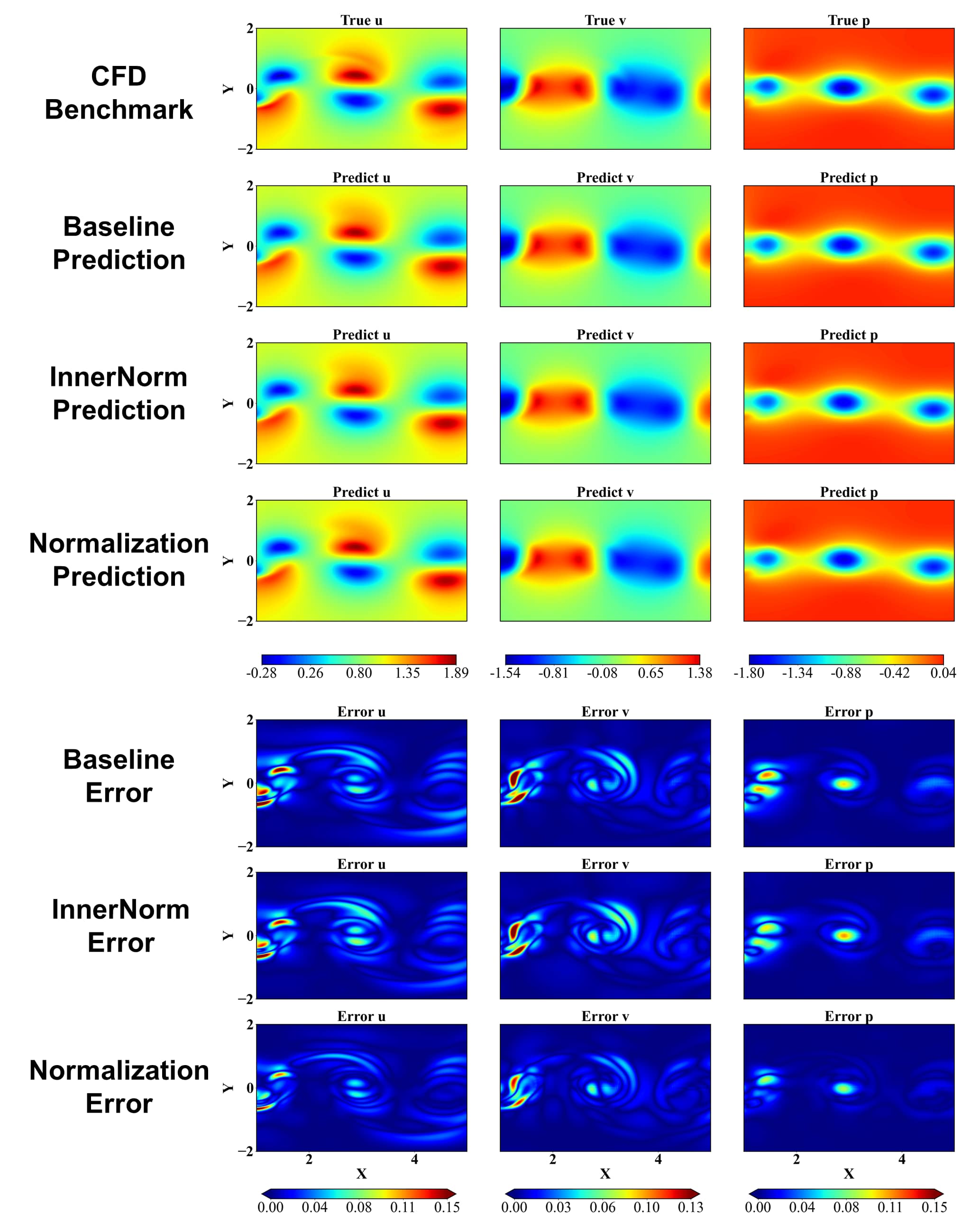}
  \caption{Flow field reconstruction result of flow past 2D circular cylinder at $\rm{Re}=10000$, at the time snapshot $t=0.58$. Contour plot and point-wise error of stream-wise velocity $u$ (first column), transverse velocity $v$ (second column) and pressure $p$ (third column). All trained with batch size of 8192, parameter size of 9731, and exponential decay learning rate.}
  \label{fig:appendixA_Re10000}
\end{figure*}
\begin{figure*}[!htbp]
  \centering
  \includegraphics[width=0.95\textwidth]{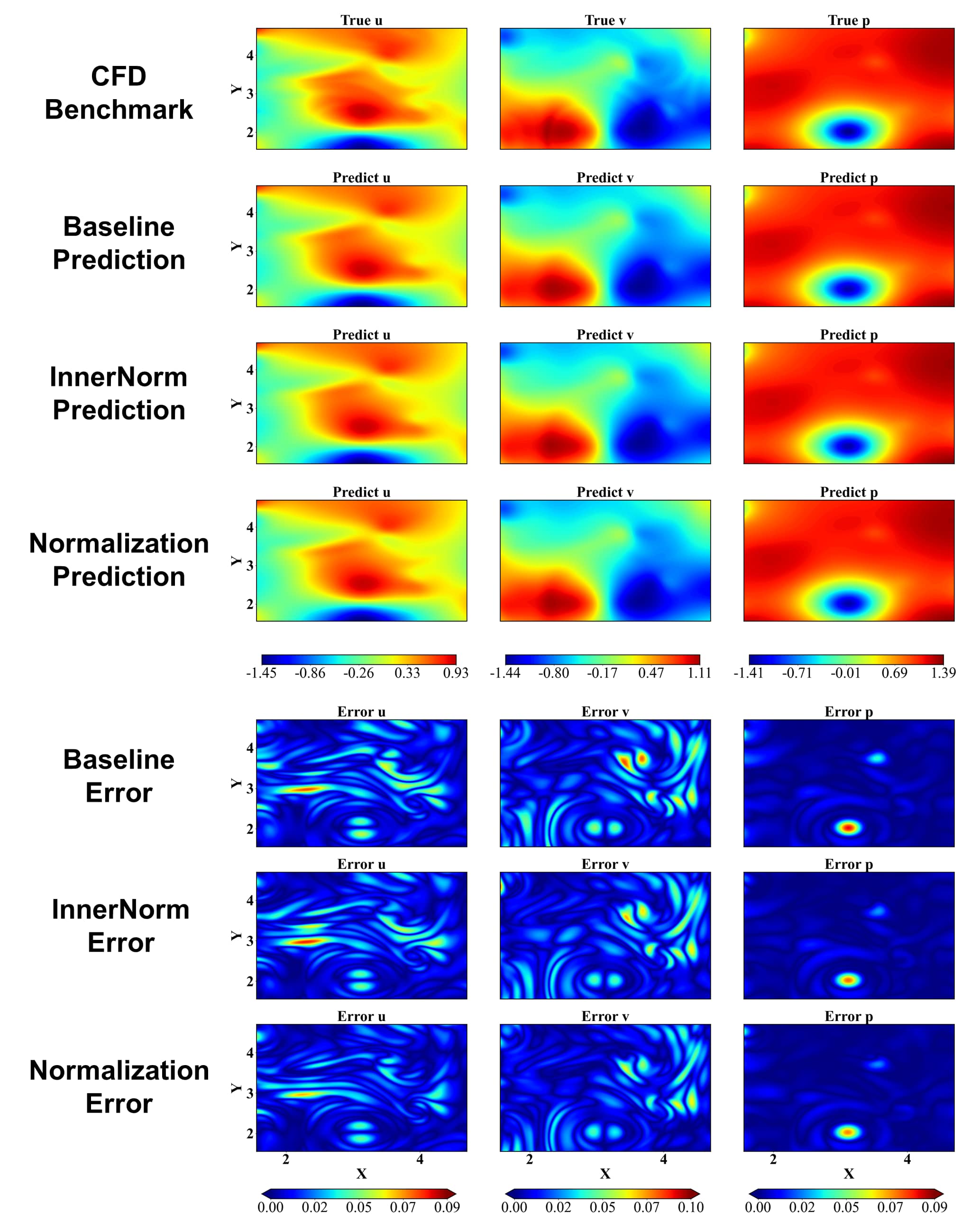}
  \caption{Flow field reconstruction result of decaying turbulence at $\rm{Re}=2000$, at the time snapshot $t=19.90$. Contour plot and point-wise error of stream-wise velocity $u$ (first column), transverse velocity $v$ (second column) and pressure $p$ (third column). All trained with batch size of 8192, parameter size of 9731, and exponential decay learning rate.}
  \label{fig:appendixA_Re2000}
\end{figure*}
\section{\label{Appendix:B}Detailed Quantification}
\renewcommand{\thefigure}{B.\arabic{figure}} 
\setcounter{figure}{0}  
\begin{figure*}[!htbp]
\centering
\subfigure[]{\label{fig:epoch_3900L2norm-u}
\includegraphics[width=0.30\linewidth]{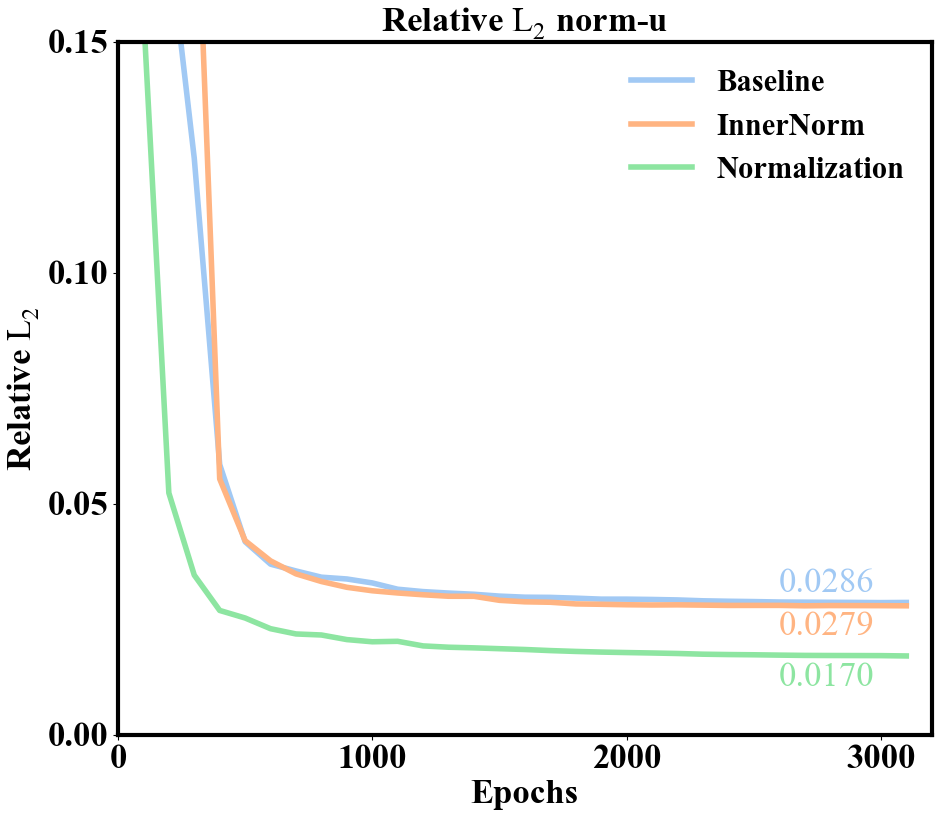}}
\hspace{0.01\linewidth}
\subfigure[]{\label{fig:epoch_3900L2norm-v}
\includegraphics[width=0.30\linewidth]{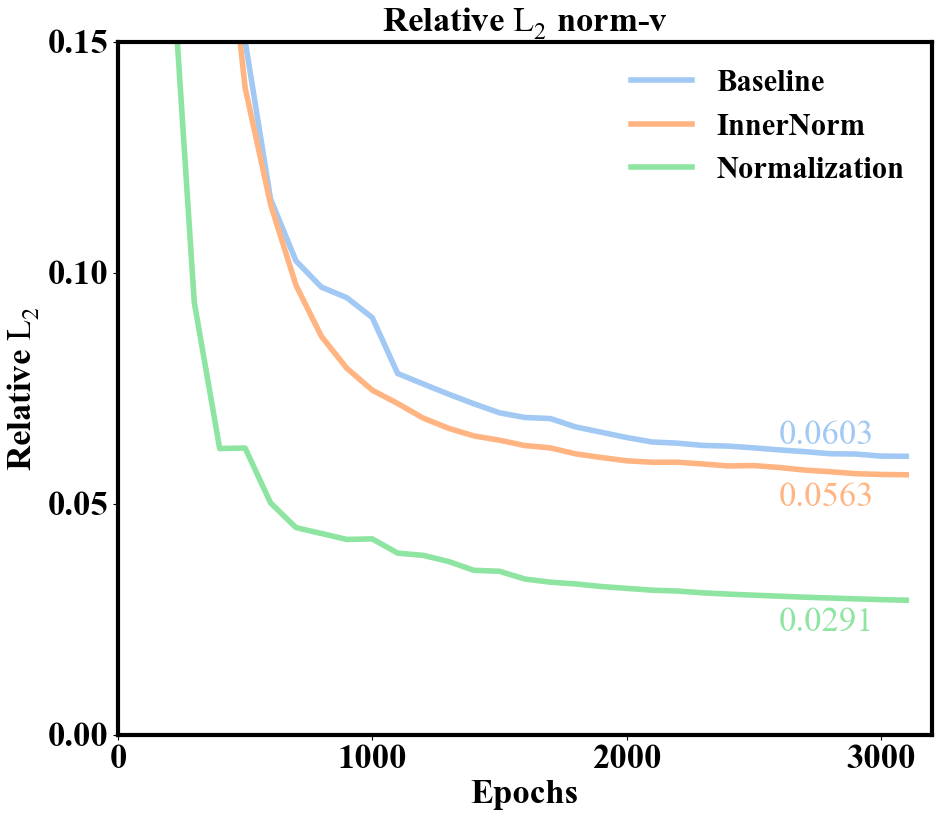}}
\hspace{0.01\linewidth}
\subfigure[]{\label{fig:epoch_3900L2norm-p}
\includegraphics[width=0.30\linewidth]{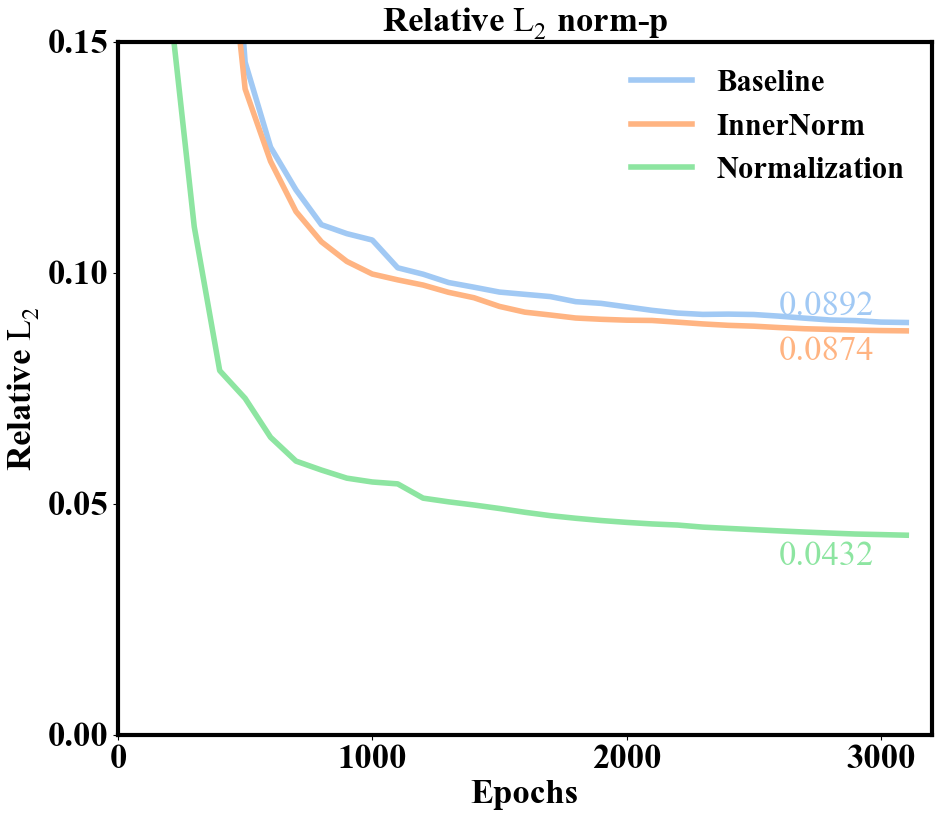}}
{\setlength{\abovecaptionskip}{0mm}
\setlength{\belowcaptionskip}{0mm}
\caption{Relative $L_{2}$ norms while training. The relative $L_{2}$ norms of (a): stream-wise velocity $u$, (b): transverse velocity $v$, (c): pressure $p$ between benchmark flow data (flow past 2D circular cylinder at $\rm{Re}=3900$) and predicted flow data . The relative L2 norms are averaged values calculated from three independent runs, all trained with batch size of 8192, parameter size of 9731, and exponential decay learning rate.}}
\label{fig:Re3900_L2norm_epoch}
\end{figure*}
\begin{figure*}[!htbp]
\centering
\subfigure[]{\label{fig:epoch_10000L2norm-u}
\includegraphics[width=0.30\linewidth]{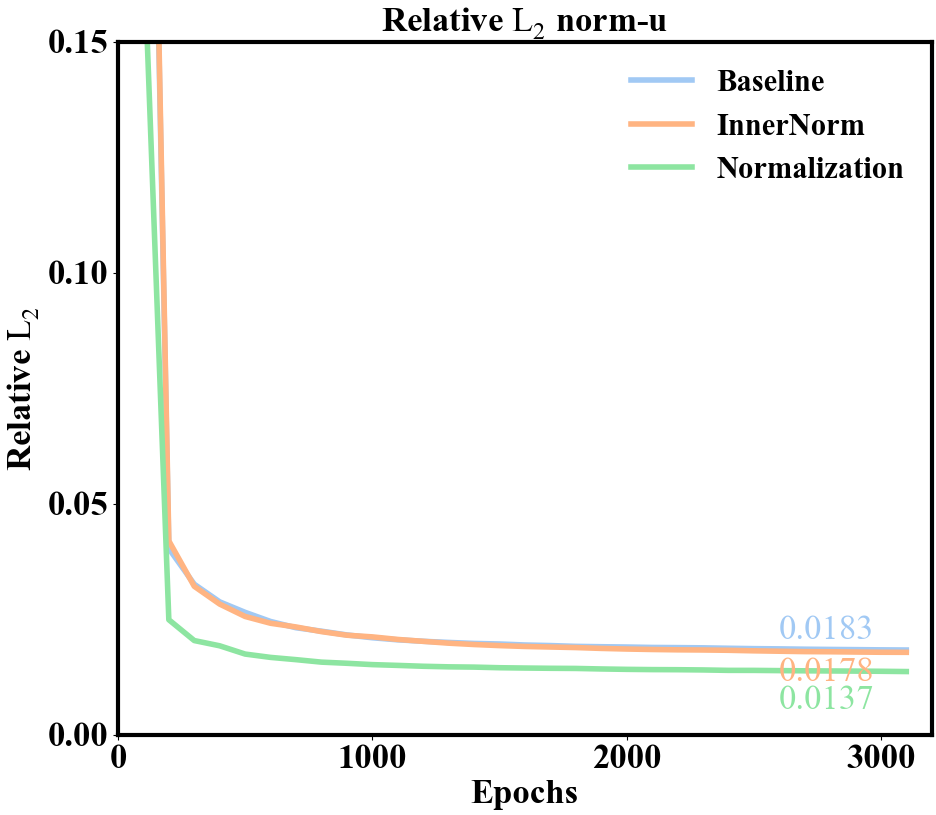}}
\hspace{0.01\linewidth}
\subfigure[]{\label{fig:epoch_10000L2norm-v}
\includegraphics[width=0.30\linewidth]{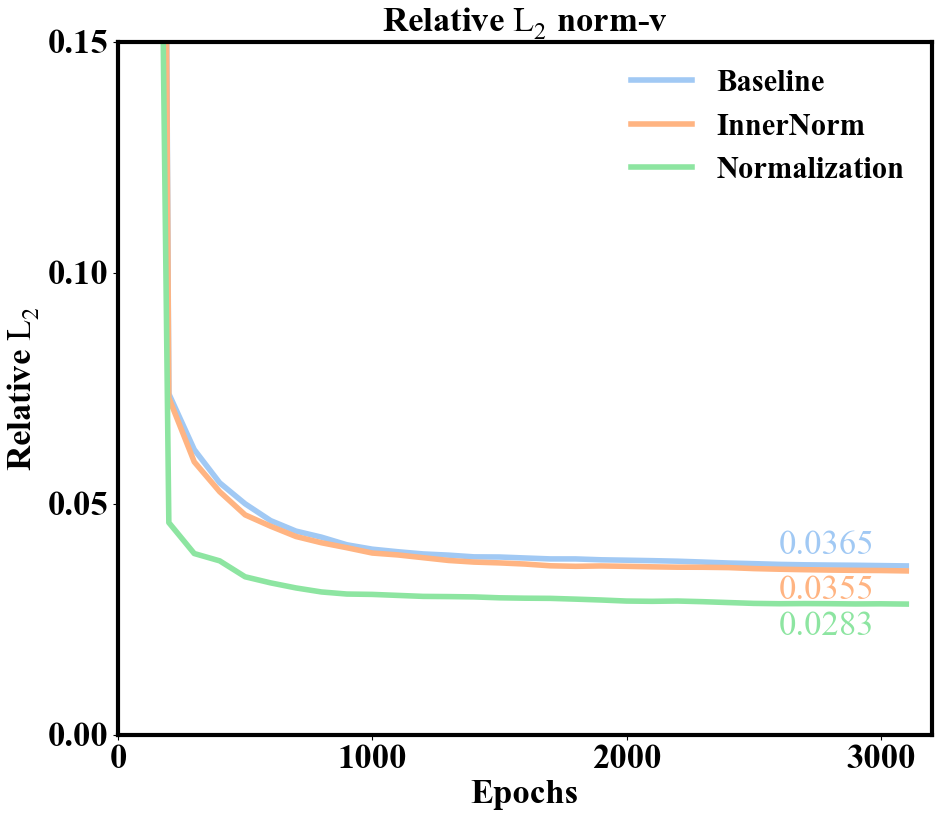}}
\hspace{0.01\linewidth}
\subfigure[]{\label{fig:epoch_10000L2norm-p}
\includegraphics[width=0.30\linewidth]{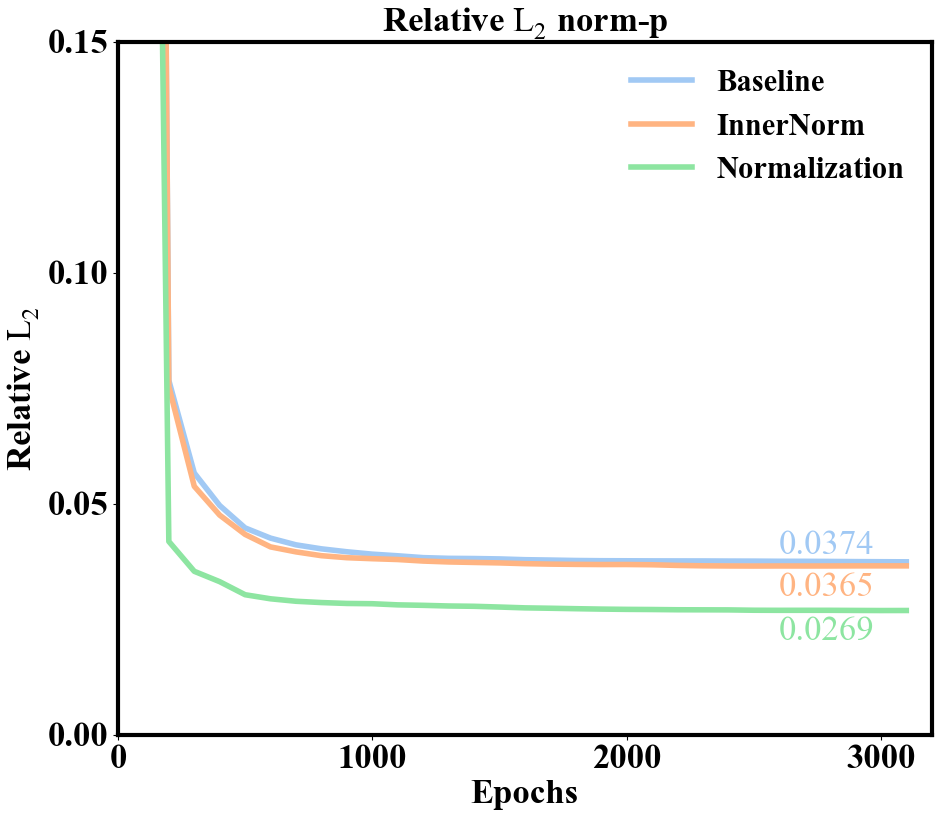}}
{\setlength{\abovecaptionskip}{0mm}
\setlength{\belowcaptionskip}{0mm}
\caption{Relative $L_{2}$ norms while training. The relative $L_{2}$ norms of (a): stream-wise velocity $u$, (b): transverse velocity $v$, (c): pressure $p$ between benchmark flow data (flow past 2D circular cylinder at $\rm{Re}=10000$) and predicted flow data. The relative L2 norms are averaged values calculated from three independent runs, all trained with batch size of 8192, parameter size of 9731, and exponential decay learning rate.}}
\label{fig:Re10000_L2norm_epoch}
\end{figure*}
\begin{figure*}[!htbp]
\centering
\subfigure[]{\label{fig:epoch_2000L2norm-u}
\includegraphics[width=0.30\linewidth]{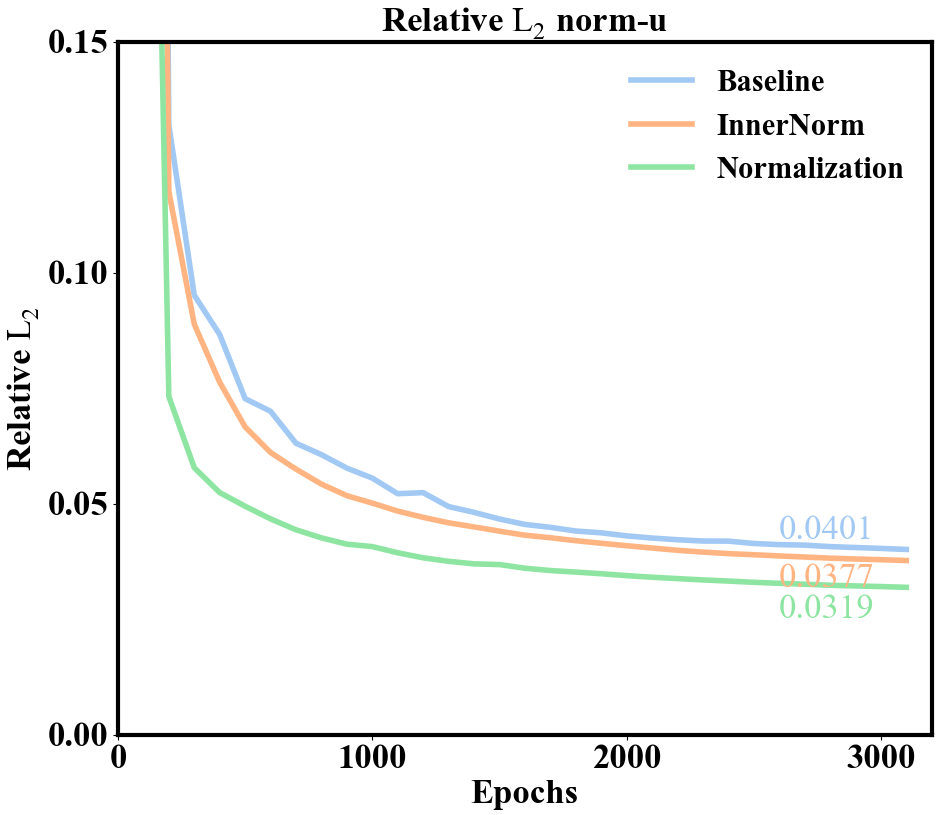}}
\hspace{0.01\linewidth}
\subfigure[]{\label{fig:epoch_2000L2norm-v}
\includegraphics[width=0.30\linewidth]{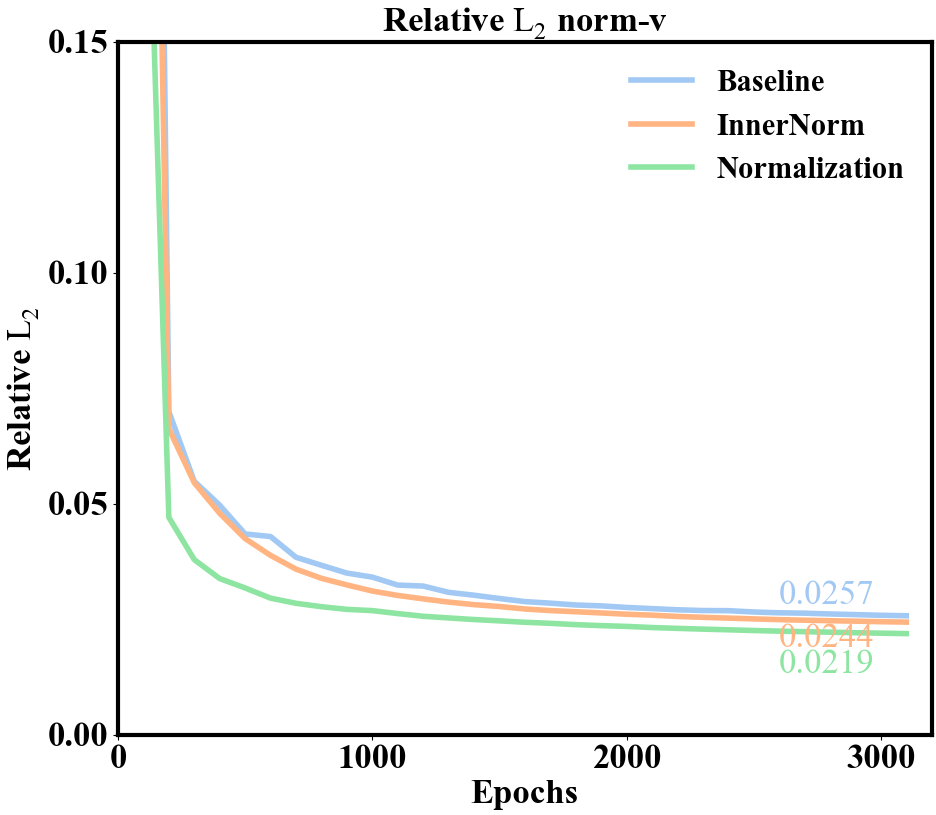}}
\hspace{0.01\linewidth}
\subfigure[]{\label{fig:epoch_2000L2norm-p}
\includegraphics[width=0.30\linewidth]{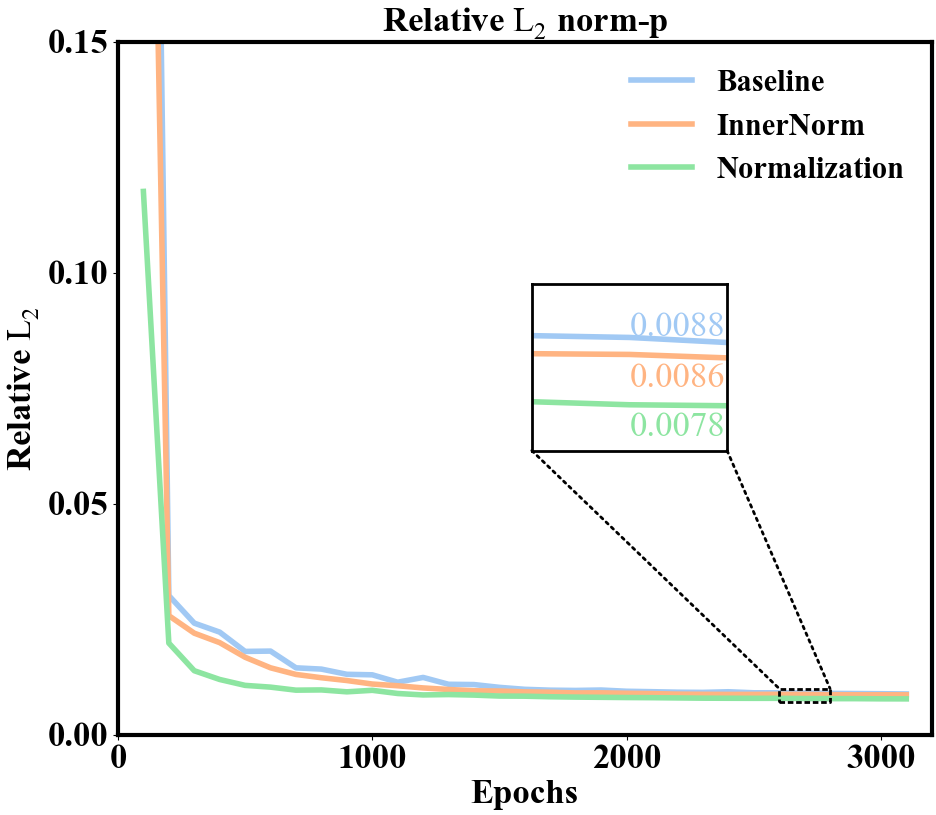}}
{\setlength{\abovecaptionskip}{0mm}
\setlength{\belowcaptionskip}{0mm}
\caption{Relative $L_{2}$ norms while training. The relative $L_{2}$ norms of (a): stream-wise velocity $u$, (b): transverse velocity $v$, (c): pressure $p$ between benchmark flow data (decaying turbulence at $\rm{Re}=2000$) and predicted flow data. The relative L2 norms are averaged values calculated from three independent runs, all trained with batch size of 8192, parameter size of 9731, and exponential decay learning rate.}}
\label{fig:Re2000_L2norm_epoch}
\end{figure*}
\begin{figure*}[!htbp]
\centering
\subfigure[]{\label{fig:final_3900L2norm-u}
\includegraphics[width=0.30\linewidth]{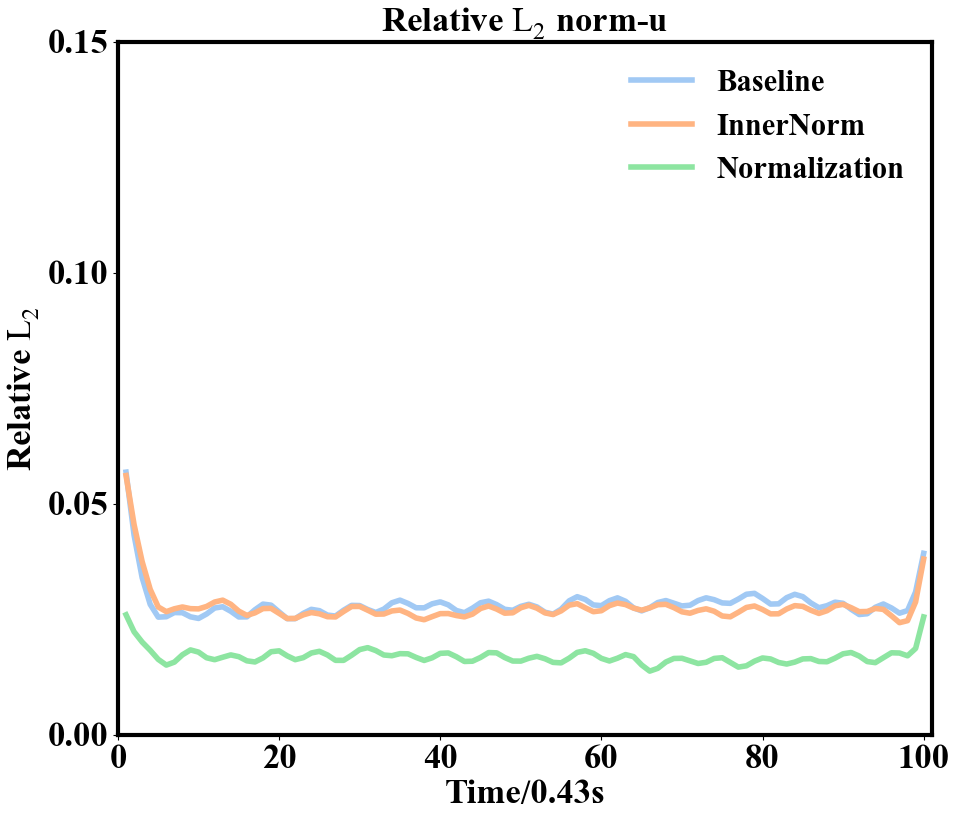}}
\hspace{0.01\linewidth}
\subfigure[]{\label{fig:final_3900L2norm-v}
\includegraphics[width=0.30\linewidth]{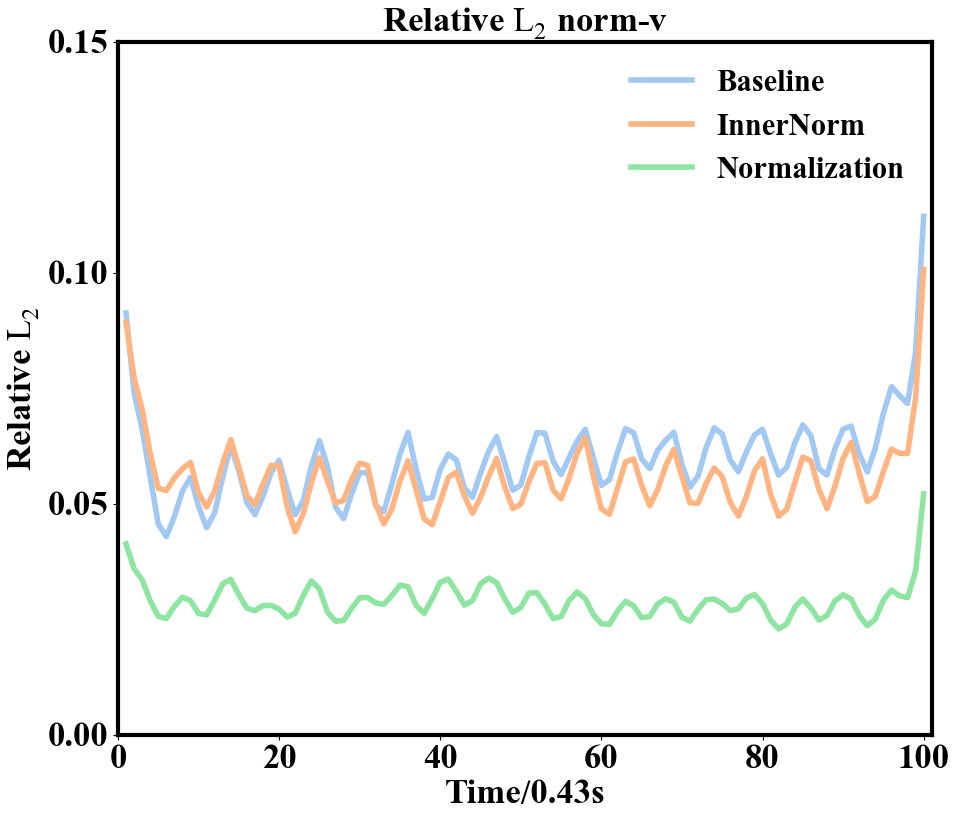}}
\hspace{0.01\linewidth}
\subfigure[]{\label{fig:final_3900L2norm-p}
\includegraphics[width=0.30\linewidth]{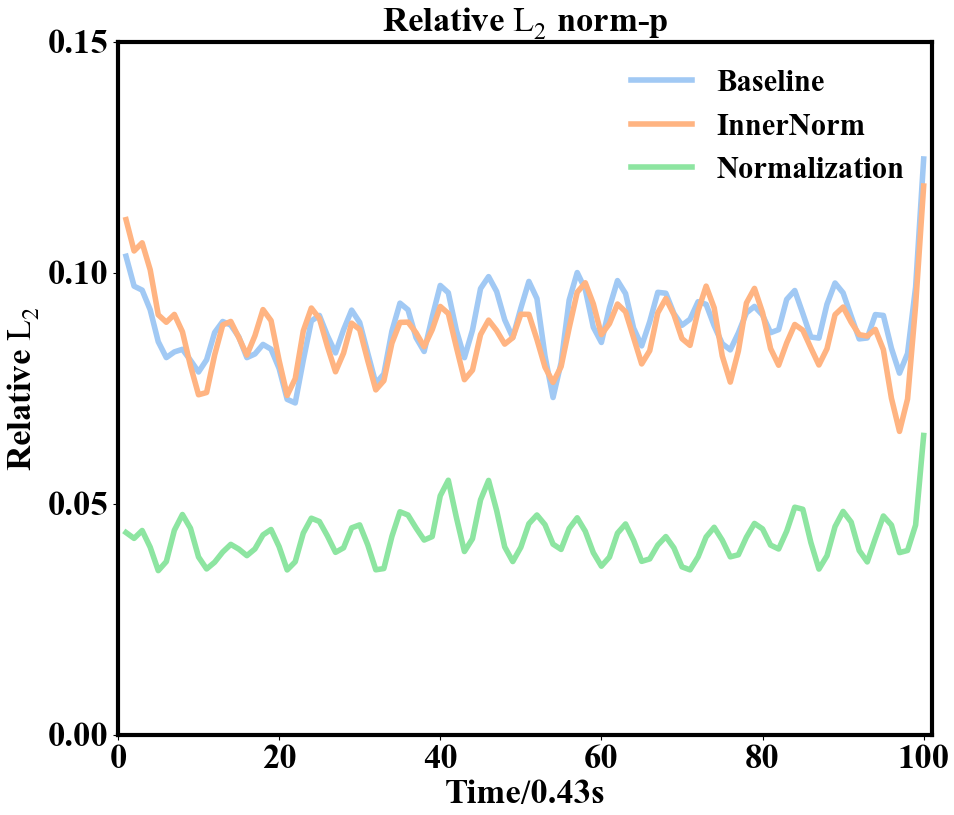}}
{\setlength{\abovecaptionskip}{0mm}
\setlength{\belowcaptionskip}{0mm}
\caption{Relative $L_{2}$ norms for each time snapshot after training. The relative $L_{2}$ norms of (a): stream-wise velocity $u$, (b): transverse velocity $v$, (c): pressure $p$ between benchmark flow data (flow past 2D circular cylinder at $\rm{Re}=3900$) and predicted flow data . The relative L2 norms are averaged values calculated from three independent runs, all trained with batch size of 8192, parameter size of 9731, and exponential decay learning rate.}}
\label{fig:Re3900_L2norm_final}
\end{figure*}
\begin{figure*}[!htbp]
\centering
\subfigure[]{\label{fig:final_10000L2norm-u}
\includegraphics[width=0.30\linewidth]{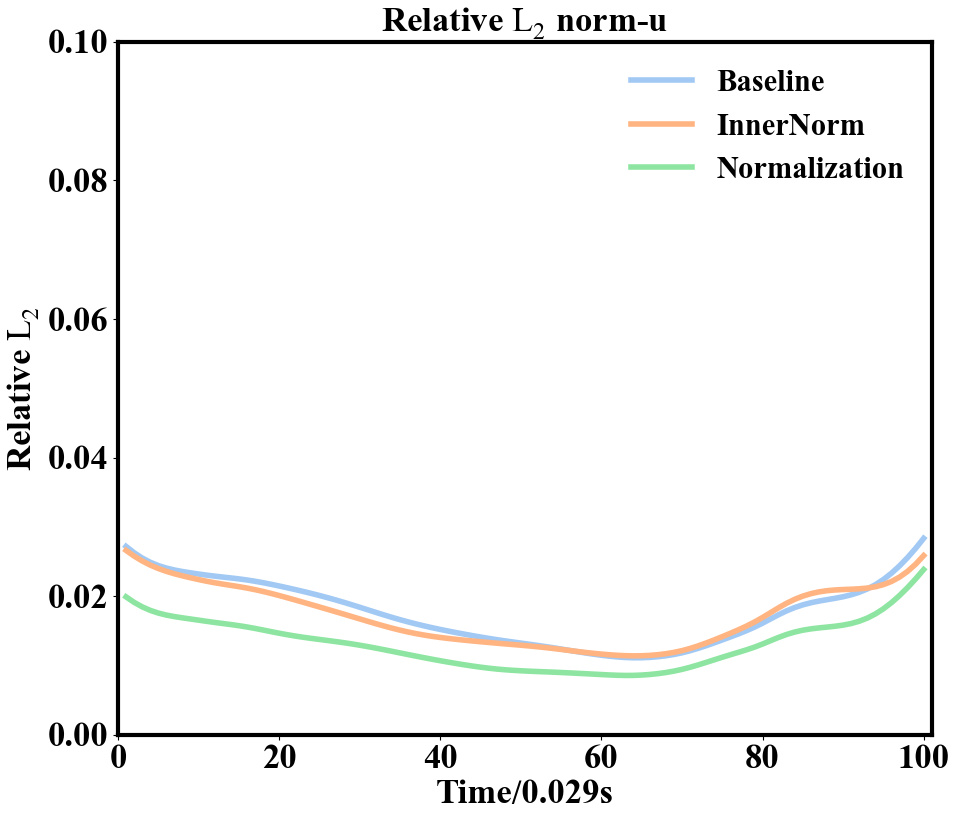}}
\hspace{0.01\linewidth}
\subfigure[]{\label{fig:final_10000L2norm-v}
\includegraphics[width=0.30\linewidth]{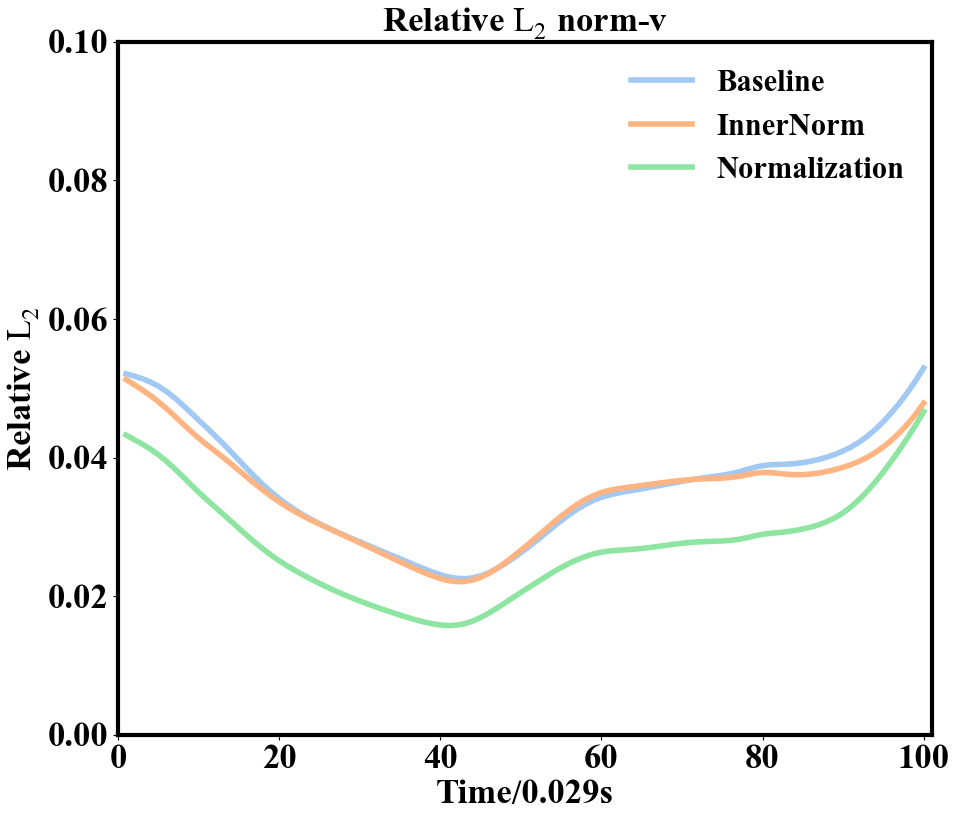}}
\hspace{0.01\linewidth}
\subfigure[]{\label{fig:final_10000L2norm-p}
\includegraphics[width=0.30\linewidth]{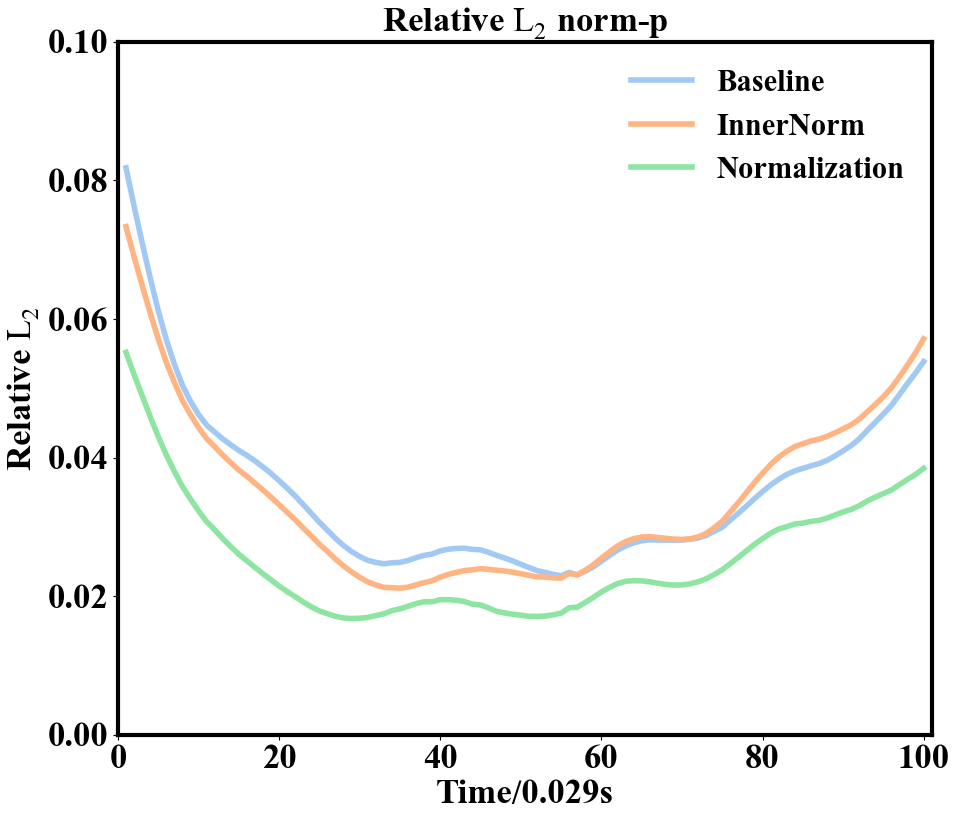}}
{\setlength{\abovecaptionskip}{0mm}
\setlength{\belowcaptionskip}{0mm}
\caption{Relative $L_{2}$ norms for each time snapshot after training. The relative $L_{2}$ norms of (a): stream-wise velocity $u$, (b): transverse velocity $v$, (c): pressure $p$ between benchmark flow data (flow past 2D circular cylinder at $\rm{Re}=10000$) and predicted flow data . The relative L2 norms are averaged values calculated from three independent runs, all trained with batch size of 8192, parameter size of 9731, and exponential decay learning rate.}}
\label{fig:Re10000_L2norm_final}
\end{figure*}
\begin{figure*}[!htbp]
\centering
\subfigure[]{\label{fig:final_20000L2norm-u}
\includegraphics[width=0.30\linewidth]{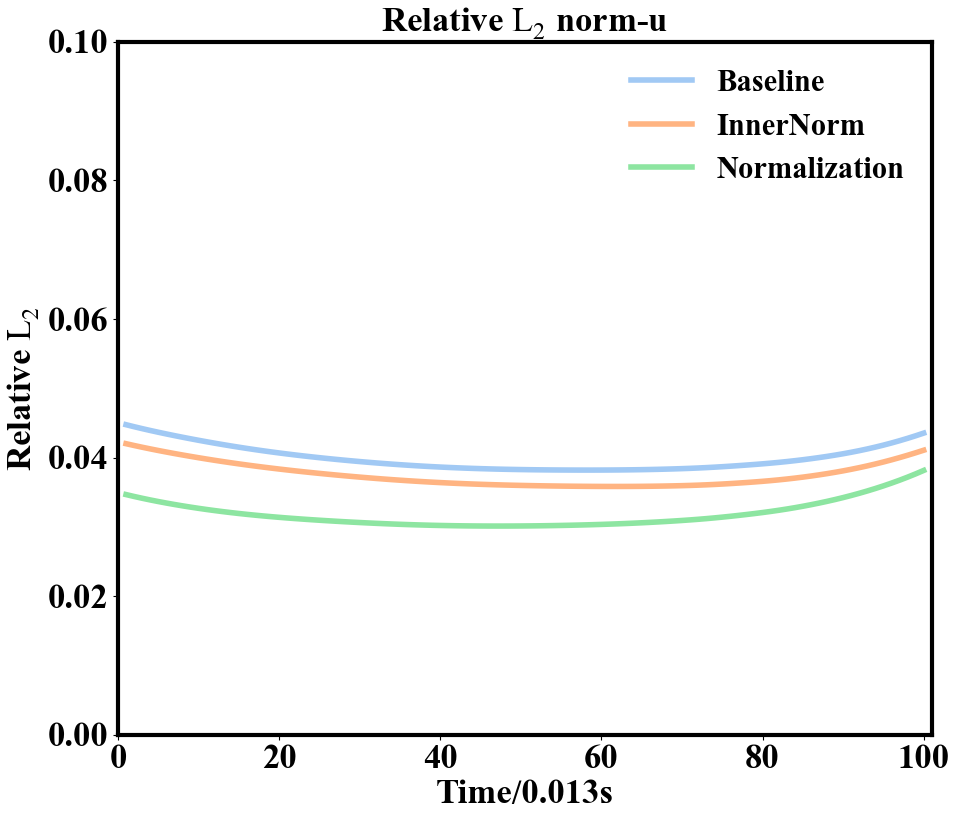}}
\hspace{0.01\linewidth}
\subfigure[]{\label{fig:final_20000L2norm-v}
\includegraphics[width=0.30\linewidth]{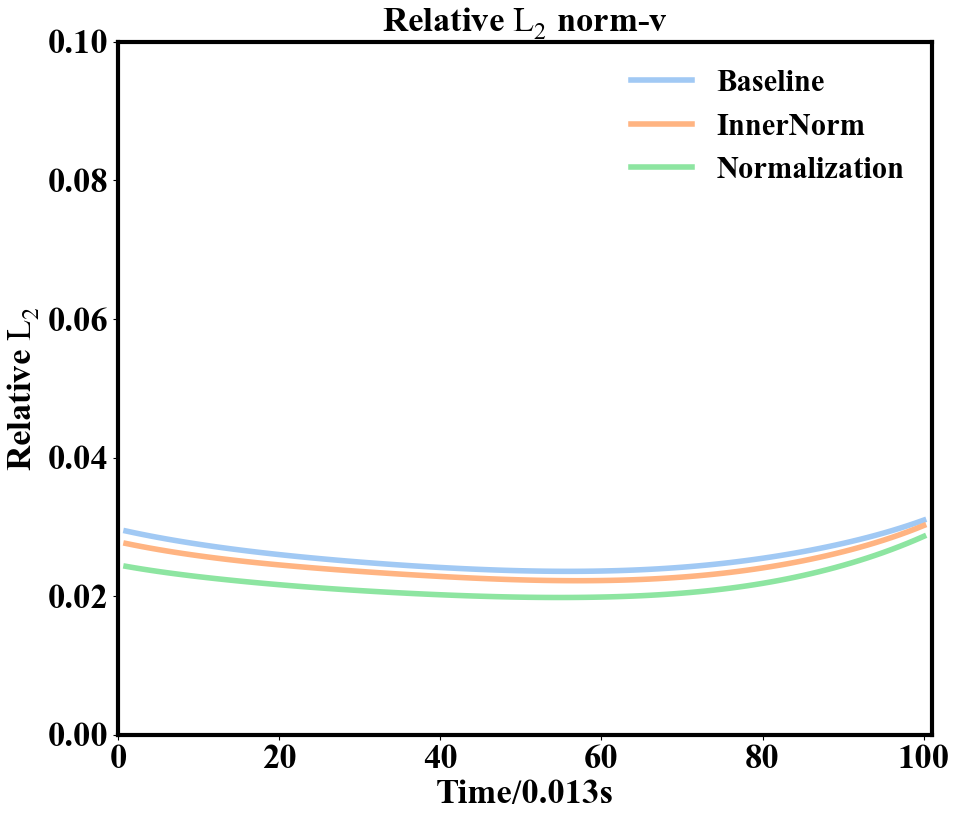}}
\hspace{0.01\linewidth}
\subfigure[]{\label{fig:final_20000L2norm-p}
\includegraphics[width=0.30\linewidth]{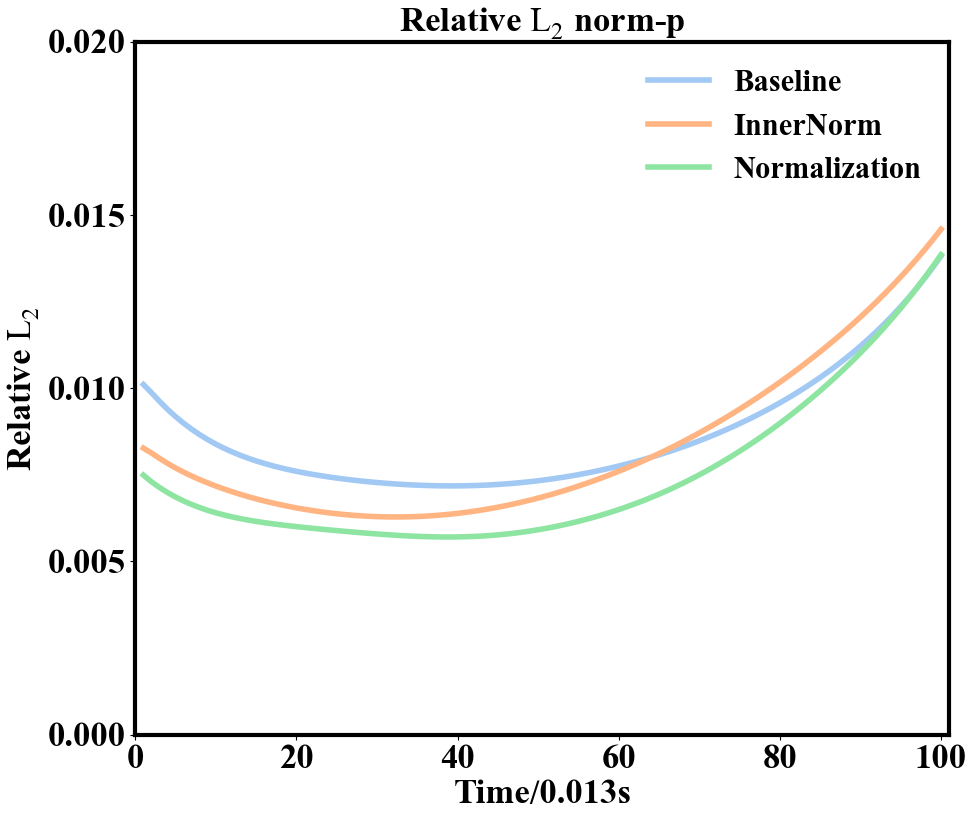}}
{\setlength{\abovecaptionskip}{0mm}
\setlength{\belowcaptionskip}{0mm}
\caption{Relative $L_{2}$ norms for each time snapshot after training. The relative $L_{2}$ norms of (a): stream-wise velocity $u$, (b): transverse velocity $v$, (c): pressure $p$ between benchmark flow data (decaying turbulence at $\rm{Re}=2000$) and predicted flow data . The relative L2 norms are averaged values calculated from three independent runs, all trained with batch size of 8192, parameter size of 9731, and exponential decay learning rate.}}
\label{fig:Re2000_L2norm_final}
\end{figure*}
\end{document}